\documentclass[fleqn,usenatbib]{mnras}
\usepackage{newtxtext,newtxmath}

\usepackage[T1]{fontenc}
\usepackage{ae,aecompl}

\usepackage{graphicx}	
\usepackage{amsmath}	
\usepackage{siunitx}
\usepackage{xspace}
\usepackage{esvect}
\usepackage{subfigure}
\usepackage{etoolbox}
 
\newcommand{\curL}{\mathcal{L}}

\newcommand{\msini}{\ensuremath{m \sin I}\xspace} 
\newcommand{\ms}{\ensuremath{\mathrm{m}\,\mathrm{s}^{-1}}\xspace}
\newcommand{\masyr}{\ensuremath{\mathrm{mas}\,\mathrm{yr}^{-1}}\xspace}
\newcommand{\kms}{\ensuremath{\mathrm{km}\,\mathrm{s}^{-1}}\xspace}

\newcommand{\Mj}{\ensuremath{M_{\rm{Jup}}}\xspace} 

\newcommand{\Rsun}{\ensuremath{R_{\odot}}\xspace}
\newcommand{\Msun}{\ensuremath{M_{\odot}}\xspace}

\title[Planet-debris disc mutual inclinations]{Mutual inclinations between giant planets and their debris discs in HD~113337 and HD~38529}

\author[]{
Jerry W. Xuan,$^{1}$\thanks{E-mail: wx239@cam.ac.uk} Grant M. Kennedy,$^{2,3}$ Mark C. Wyatt,$^{1}$ and Ben Yelverton$^{1}$
\\
$^{1}$Institute of Astronomy, University of Cambridge, Madingley Road, Cambridge CB3 0HA, UK \\
$^{2}$Department of Physics, University of Warwick, Gibbet Hill Road, Coventry, CV4 7AL, UK\\
$^{3}$Centre for Exoplanets and Habitability, University of Warwick, Gibbet Hill Road, Coventry CV4 7AL, UK\\
}

\date{Accepted 2020 October 1. Received 2020 August 14; in original form 2020 August 14}

\pubyear{2020}

\begin{document}
\label{firstpage}
\pagerange{\pageref{firstpage}--\pageref{lastpage}}
\maketitle

\begin{abstract}

HD~113337 and HD~38529 host pairs of giant planets, a debris disc, and wide M-type stellar companions. We measure the disc orientation with resolved images from {\it Herschel} and constrain the three-dimensional orbits of the outer planets with {\it Gaia} DR2 and {\it Hipparcos} astrometry. Resolved disc modelling leaves degeneracy in the disc orientation, so we derive four separate planet-disc mutual inclination ($\Delta I$) solutions. The most aligned solutions give $\Delta I=17-32\degr$ for HD~113337 and $\Delta I=21-45\degr$ for HD~38529 (both $1\sigma$). In both systems, there is a small probability ($<0.3$ per cent) that the planet and disc are nearly aligned ($\Delta I<3\degr$). The stellar and planetary companions cause the orbits of disc material to precess about a plane defined by the forced inclination. We determine this as well as the precession time-scale to interpret the mutual inclination results. We find that the debris discs in both systems could be warped via joint influences of the outer planet and stellar companion, potentially explaining the observed misalignments. However, this requires HD 113337 to be old (0.8-1.7 Gyr), whereas if young (14-21 Myr), the observed misalignment in HD 113337 could be inherited from the protoplanetary disc phase. For both systems, the inclination of the stellar spin axis is consistent with the disc and outer planet inclinations, which instead supports system-wide alignment or near alignment. High-resolution observations of the discs and improved constraints on the planetary orbits would provide firmer conclusions about the (mis)alignment status.

\end{abstract}


\section{Introduction}\label{sec:intro}
Debris discs are thought to be the dusty remnants of gaseous protoplanetary discs, and extra-solar analogues to the Kuiper belt \citep[see reviews by][]{wyatt_evolution_2008, hughes_debris_2018}. With far-infrared instruments such as {\it Spitzer} and {\it Herschel}, debris discs are detected around $\sim 20$ per cent of FGK stars \citep{bryden_frequency_2006, trilling_debris_2008, montesinos_incidence_2016, sibthorpe_analysis_2018}. Debris discs may be a by-product of planet formation, and a number of systems are found to host both debris discs and planets, allowing rich studies of planet-disc interactions. For example, direct imaging has revealed giant planets orbiting $\beta$ Pic \citep{lagrange_probable_2009-1, lagrange_giant_2010-1} and HR 8799 \citep{marois_direct_2008-2, marois_images_2010}, stars with bright debris discs.

The mutual inclination ($\Delta I$) between the debris disc plane and a planet's orbital plane is a key parameter in understanding the dynamical history of a planetary system. As with measurements of mutual inclination between planets \citep[e.g.][]{mcarthur_new_2010, mills_kepler-108_2017, xuan_evidence_2020}, measurements of the planet-disc mutual inclination could have implications for the early conditions of planetary formation. In the solar system, for example, the orbital planes of planets are closely aligned with each other, as well as to the Kuiper belt, suggesting a picture of planet formation in a flat protoplanetary disc. 

Previous studies have also revealed several systems consistent with alignment or near alignment between debris discs and planets. $\beta$ Pic b's orbit is found to be nearly aligned with the disc mid-plane with $\Delta I\approx2.4\degr$ \citep{matra_kuiper_2019}, and also closely aligned with the stellar spin axis \citep{kraus_spin-orbit_2020-1} and the orbit of the inner planet \citep{nowak_direct_2020}. In HR 8799 \citep{matthews_resolved_2014} and HD 82943 \citep{kennedy_star_2013}, the planets are consistent with being aligned with the debris disc and the stellar spin axis, as inferred by similarities in their sky-projected inclinations. Recently, AU Mic was found to host a short-period transiting planet that is aligned with its debris disc and stellar spin axis \citep{plavchan_planet_2020, palle_transmission_2020}. Depending on the system age, such alignments could either be primordial, or the result of subsequent dynamical interactions. For example, secular perturbations from companions would make the orbits of planetesimals in the disc precess, resulting in a disc mid-plane that could differ from its initial plane. As a result, a disc that was initially misaligned with a planet can become aligned with it given time for a few precession periods to occur, although the disc would be reshaped in the process \citep[e.g.][]{mouillet_planet_1997, kennedy_99_2012, pearce_dynamical_2014}.

Strictly speaking, the sky-projected inclination ($I$) and longitude of ascending node ($\Omega$) of both the debris disc and planetary orbit are needed to measure the planet-disc mutual inclination, but a planet's inclination and ascending node cannot usually be constrained without astrometric measurements. Some of the past studies (e.g. for HD 82943, AU Mic) do not have access to the planet's ascending node, and therefore used the fact that the planet and disc inclinations are consistent to conclude alignment. Although conclusions of alignment from similar inclination values are relatively robust on probabilistic grounds (especially if the stellar inclination also agrees with the planet and disc inclinations), such an approach could miss truly misaligned systems, because differing values of $\Omega$ between the disc and planet result in non-zero mutual inclinations. 

Despite several systems that are found to have aligned or nearly aligned debris discs and planets, the number of measurements remain low, and there could be as yet undetected, misaligned systems. Recently, \citet{xuan_evidence_2020} found $\Delta I=49-131\degr$ ($1\sigma$) between the inner super-Earth and outer giant planet in $\pi$ Men (see similar results from \citealt{damasso_precise_2020-1}, \citealt{de_rosa_significant_2020-1}), which is known to host a debris disc from infrared excess emission \citep{sibthorpe_analysis_2018}. As argued in \citet{xuan_evidence_2020}, imaging the disc and measuring the mutual inclination between the disc and giant planet in $\pi$ Men would provide strong constraints on how the large planet-planet $\Delta I$ arose, and how planet formation proceeded in this system. In contrast to flatter systems like the solar system, systems like $\pi$ Men could have seen a more violent dynamical history.

From the point of view of protoplanetary discs, there is growing evidence of primordial misalignments between inner and outer components of the disc. These misaligned protoplanetary discs have been observed in several systems, with mutual inclinations between the inner and outer discs ranging from $30-80\degr$ \citep[e.g.][]{marino_shadows_2015, loomis_multi-ringed_2017, min_connecting_2017, walsh_co_2017}. In these systems, the inner discs are typically located between sub-au to a few au in distance, and outer discs are located between dozens to hundreds of au. Theoretical work has shown that misalignments could result from misaligned stellar or planetary-mass companions within the system \citep{zhu_inclined_2019,nealon_scattered_2019}. We might expect planetary systems that emerge from such discs to show observable signatures such as misalignments between the descendant debris disc and planets.

In this paper, we study planet-disc alignment in HD~113337 and HD~38529, which both host multiple giant planets \citep{fischer_planetary_2003, borgniet_extrasolar_2019}, debris discs \citep{moro-martin_dust_2007, rhee_characterization_2007}, and wide M-type stellar companions \citep{montes_calibrating_2018}. In both systems, the outer planets detected by radial velocity (RV) have masses ($m$) and semimajor axes ($a$) comparable to directly imaged planets like $\beta$ Pic b ($m\sim13~\Mj$ and $a\sim11$ au, \citealt{gravity_collaboration_peering_2020}): HD~113337~c has $\msini\sim7~\Mj$ and $a\sim4.8$ au \citep{borgniet_extrasolar_2019}, while HD~38529~c has $\msini\sim13~\Mj$ and $a\sim3.6$ au \citep{fischer_planetary_2003}. In these two systems, comparison of {\it Gaia} DR2 and {\it Hipparcos} astrometry shows that the stars exhibit significant differences in proper motion indicative of orbital motion \citep[e.g][]{brandt_hipparcos-gaia_2018, kervella_stellar_2019}, which we find enables full orbit characterizations of the outer planets. Furthermore, the debris discs of HD~113337 and HD~38529 are resolved by {\it Herschel} at 70~$\mu$m. The combination of these two observational factors renders the systems ideal tests for planet-disc alignment.

To determine the full orbit (including the inclination and longitude of ascending node) of the outer planets, we jointly fit {\it Gaia} DR2 and {\it Hipparcos} astrometry with literature RV data. Then, we measure the orientations of the debris discs using the {\it Herschel} data. By combining these measurements, we directly compute the mutual inclinations between the outer planets and their exterior discs. For both systems, the most aligned solutions give median values of $\Delta I\sim25-30\degr$ between the planet and disc, and alignment is ruled out with $\gtrsim3\sigma$ confidence ($<0.3$ per cent chance). To interpret these results, we study the dynamics of orbiting planetesimals in the disc under the influence of the outer planets and stellar binaries, which we find to be important for the evolution of the disc.

We organize this paper as follows. In \S\ref{sec:113337_intro} and \S\ref{sec:38529_intro}, we introduce the star, planetary system, and debris disc for HD~113337 and HD~38529 respectively. Then, we present measurements of the debris disc orientations in \S\ref{sec:disc_model}. In \S\ref{sec:companion_orbits}, we describe our joint astrometric and RV fits to the orbits of HD~38529~c and HD~113337 c, the outer planets in the systems. We discuss the alignment status of the planets and debris discs in \S\ref{sec:assess_align}, before discussing our results in \S\ref{sec:discuss}.

\section{The HD~113337 system}\label{sec:113337_intro}
\subsection{The star}
HD~113337 is a main-sequence F6V star at a distance of $36.2\pm0.1$pc, based on the Gaia~DR2 parallax measurement \citep{gaia_collaboration_2018}. As a field star, the stellar age is difficult to constrain. Based on interferometric measurements of the star's angular diameter, \citet{borgniet_constraints_2019} measured $R_\star=1.50\pm0.04\Rsun$ and found two distinct solutions for the system age, one young within 14-21 Myr (with $M=1.48\pm0.08\Msun$) and one old between 0.8-1.7 Gyr (with $M=1.40\pm0.04\Msun$). In terms of stellar rotation, \citet{borgniet_extrasolar_2014} reported $v \sin{I_\star} = 6.3\pm1~\kms$. For the rotation period, \citet{borgniet_extrasolar_2019} found strong peaks of $\sim2$ and $\sim4$\,d in the Lomb Scargle periodogram of the RV residuals. In addition, they found $\sim2$ and $\sim3$\,d periodicities in the bisector velocity span, which was used as a diagnosis of the stellar variability. These signals all have false alarm probabilities of $<1$ per cent, and are attributed to stellar rotation by \citet{borgniet_extrasolar_2019}. To estimate the stellar spin inclination, we assume the $\sim2-4$ d periods are indeed close to the stellar rotation period, and adopt $P_\star=3\pm1$\,d. Combining $R_\star$, $v \sin{I_\star}$, and $P_\star$, we get $I_\star=24^{+30}_{-10}\degr$, which will be compared to the inclinations of the outer planet and debris discs in \S\ref{sec:stellar_spin}.

We note that HD~113337 has a wide M-type stellar companion (2MASS J13013268+6337496) with a projected separation of $\sim120$ arcsec (or $\sim4000$ au), and a position angle of $\sim307\degr$ \citep{reid_meeting_2007}. The stellar companion is confirmed from common proper motion and parallax \citep{montes_calibrating_2018}, and is itself found to be a binary pair of M3.5V stars from high-resolution imaging, with estimated masses of $\sim0.25\Msun$ for each component and a total mass of $\sim0.5\Msun$ \citep{janson_astralux_2012}. 

\subsection{The giant planets}
HD~113337 hosts one confirmed giant planet (b, $P\sim323$ d, $\msini\sim3~\Mj$) and one candidate giant planet (c, $P\sim3265$ d, $\msini\sim7~\Mj$), which were both discovered using the SOPHIE echelle fibre-fed spectrograph at the Haute Provence Observatory \citep{bouchy_sophie_2006} and reported by \citet{borgniet_extrasolar_2014} and \citet{borgniet_extrasolar_2019}. The outer planet was reported as a candidate because the measured orbital period of 3265 d is just below the SOPHIE observation data time span of 3368 d and long-term variations are visible in the activity indicators. However, \citet{borgniet_extrasolar_2019} argued that HD~113337~c is most likely real because there exists a substantial phase shift between the long-term variations in RV and that seen in the activity indicators. After removing a Keplerian model of HD~113337~b, the semi-amplitude of the RV residuals ($\sim 80\ms$) is also larger than that expected from stellar-activity induced variation \citep{lovis_harps_2011}. In this paper, we find that HD~113337 exhibits a proper motion anomaly that is consistent in size with the reflex motion that HD~113337~c would induce on its host star, based on its RV measured parameters. This strengthens the case for HD~113337~c as a real planet, which we assume in this paper.

\subsection{The debris disc}
HD~113337 shows infrared excess emission from about 20~$\mu$m up to 1.2 mm, with a fractional luminosity of \SI{1e-4}, which is inferred to arise from a debris disc \citep{rhee_characterization_2007, moor_structure_2011, chen_spitzer_2014}. \citet{borgniet_constraints_2019} modelled the outer disc as resolved by the {\it Herschel} Photodetector Array Camera and Spectrograph (PACS) at 70~$\mu$m (\texttt{OT2\_ksu\_3}), and measured a disc inclination of $I_{\rm{disc}}=25^{+5}_{-15}\degr$, a position angle of $128\pm5\degr$, and a size of $85\pm20$ au. To ensure our modelling is consistent between HD~113337 and HD~38529, we reanalyse the disc in \S\ref{sec:disc_model} using the same {\it Herschel} data to measure its orientation and size.

\begin{figure*}
\centering
\begin{subfigure}
    \centering
    \includegraphics[width=1.0\linewidth]{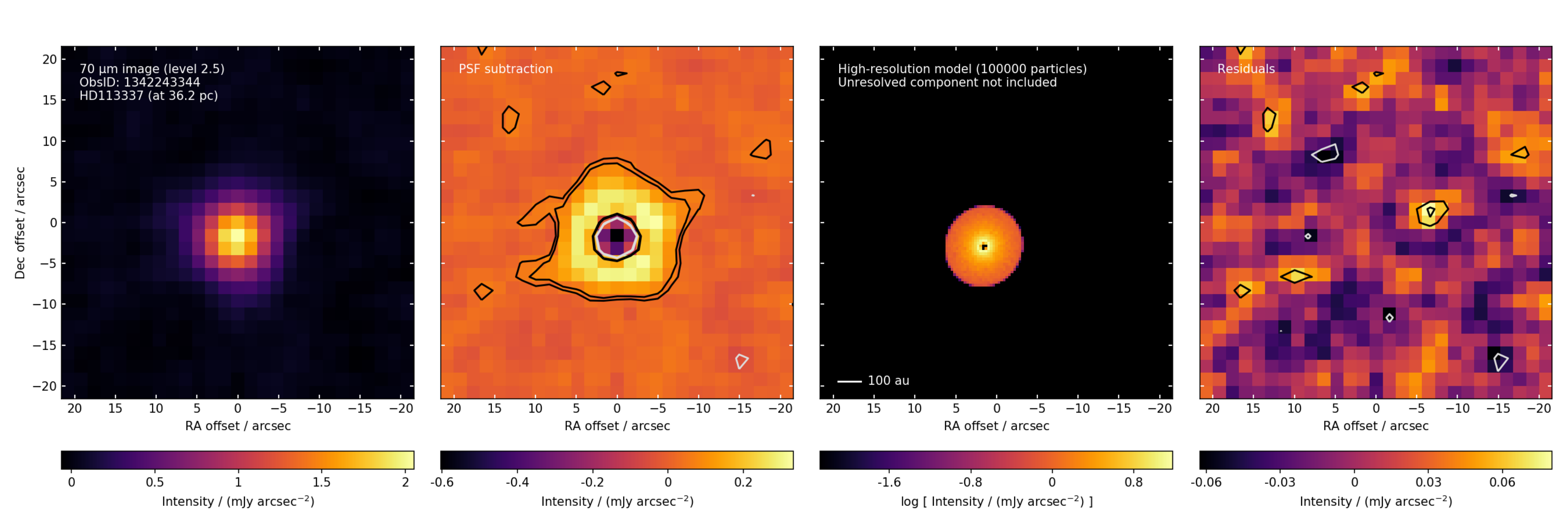}
\end{subfigure}
\begin{subfigure}
    \centering
    \includegraphics[width=1.0\linewidth]{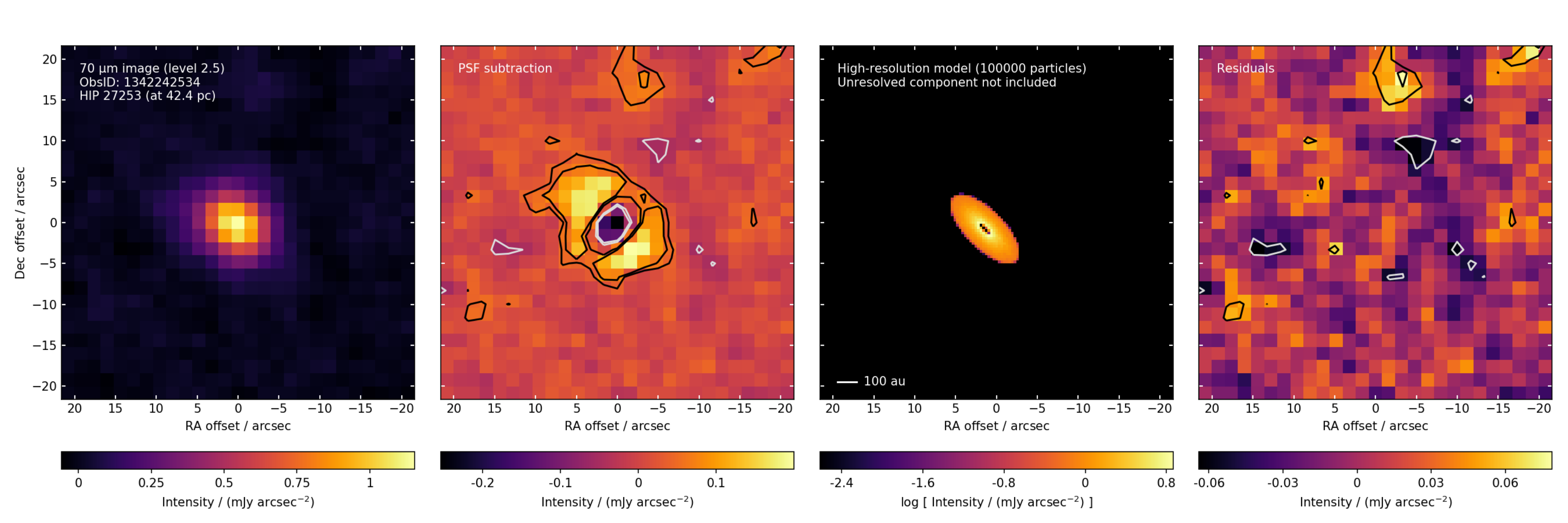}
\end{subfigure}
    \caption{Resolved disc fitting results for HD~113337 (upper row) and HD~38529 (lower row). From left to right the panels show i) data, ii) PSF-subtracted image, where the PSF was scaled to the peak pixel and allowed to have a variable offset chosen to minimize the $\chi^2$ of the PSF subtraction, iii) high-resolution image of best-fit disc model, and iv) residuals, with contours shown at $\pm$2, 3$\sigma$. In both cases the disc model is a good fit to the data.}
    \label{fig:disc_images}
\end{figure*}

\section{The HD~38529 system}\label{sec:38529_intro}
\subsection{The star}
HD~38529 is a G4IV subgiant at a distance of $42.4\pm0.1$ pc, based on the Gaia~DR2 parallax \citep{gaia_collaboration_2018}. From a direct measurement of the stellar diameter, the stellar radius is found to be $R_\star=2.58\pm0.08\Rsun$, with a derived mass of $1.36\pm0.02\Msun$ and age of $4.5\pm0.2$ Gyr \citep{henry_host_2013}. Measurements of the stellar rotation period $P_\star$ include 35.7 d \citep{fischer_planetary_2003}, $31.7\pm0.2$ d \citep{benedict_mass_2010} and $37.0\pm0.4$ d \citep{henry_host_2013}, while projected rotation speeds are found be to $v \sin{I_\star} = 3.2\pm0.5\kms$ \citep{henry_host_2013} and $3.5\pm0.5\kms$ \citep{fischer_planetary_2003}. We adopt values for $R_\star$, $P_\star$, and $v \sin{I_\star}$ from \citep{henry_host_2013}, which had more observations to determine $P_\star$ and also took advantage of a direct stellar radius measurement. This gives $I_\star=69\pm14\degr$, which we will use in \S\ref{sec:stellar_spin}.

HD~38529 also has a wide M-type stellar companion identified by \citet{raghavan_two_2006-1}. The M2.5V companion has been confirmed with common proper motion and parallax, and has a projected separation of $\sim284$ arcsec (or $\sim11000$ au), and a position angle of $\sim305\degr$ \citep{montes_calibrating_2018}. Based on the spectral type, we estimate a mass of $\sim0.35\Msun$ for the stellar companion using spectral type-mass relations from \citet{kraus_stellar_2007}.

We find that HD~38529~B has {\it Gaia} DR2 position and proper motion measurements, and attempt to constrain its orbit using the Linear Orbits for the Impatient algorithm (LOFTI) \citep{pearce_orbital_2020},\footnote{The code is freely available at \url{https://github.com/logan-pearce/LOFTI}.} which is a modified version of the rejection-sampling methodology described in \citet{blunt_orbits_2017-2}. For the HD~38529 binary pair, there are four available constraints: the relative positions in RA and Dec ($\Delta\alpha$, $\Delta\delta$) and the relative proper motions in RA and Dec ($\Delta\mu\alpha$, $\Delta\mu\delta$), all computed from {\it Gaia} DR2 astrometry. Specifically, $\Delta\mu\alpha=-0.178\pm0.024$ kms$^{-1}$ and $\Delta\mu\delta=-0.013\pm0.024$ kms$^{-1}$, so the relative velocity between the two stars has a $\sim7.4\sigma$ significance. However, as only four out of six required degrees of freedom are available (missing the relative radial velocity and distance) at a single epoch, the problem is under-constrained and the derived solutions are highly dependent on the assumed priors \citep{pearce_constraining_2015}. In LOFTI, uniform priors are adopted for eccentricity $e$, cosine of inclination ($\cos{I}$), argument of periastron ($\omega$), and mean anomaly from which to calculate the time of periastron ($T_p$), while the semimajor axis $a$ and longitude of ascending node $\Omega$ are chosen to match the $\Delta\alpha$ and $\Delta\delta$ values \citep{pearce_orbital_2020}.

Before running LOFTI, we first check that HD~38529~B can be gravitationally bound to HD~38529 using the parameter $B$ defined in \citet{pearce_constraining_2015}, which is the squared ratio between the instantaneous relative velocity and escape velocity at the projected separation. We get $B\sim0.13<1$ for HD~38529~B, indicating it is consistent with being bound. However, this low value of $B$ means that the orbital parameters are poorly unconstrained by observations over short orbital arcs (in this case a single epoch) \citep{pearce_constraining_2015}. Rather, only the allowed ranges of fitted orbital parameters can be taken with confidence, as these ranges are prior-independent \citep{pearce_constraining_2015}. Assuming the priors stated above, we ran LOFTI on the relative {\it Gaia} measurements until 200,000 orbits were accepted, and plot the joint posterior distributions in Fig.~\ref{fig:38529B_orbit}. The accepted orbits have off-by-$\pi$ degeneracies in $\omega$ and $\Omega$, which is expected from the lack of constraints in the radial direction \citep{pearce_constraining_2015, pearce_orbital_2020}. As shown in Fig.~\ref{fig:38529B_orbit}, allowed values for $I$ range widely from $91-180\degr$ (i.e. the orbital direction is clockwise as seen by the observer), with a $2\sigma$ range of $97-159\degr$. All values of $e$ are possible. In \S\ref{sec:deltaI}, we estimate possible ranges for the mutual inclination between HD~38529~B and HD~38529~c based on these results. We note that the same method is not available for HD~113337, whose binary pair of M dwarfs lack {\it Gaia} DR2 measurements.

\subsection{The giant planets}
Like HD~113337, HD~38529 also hosts two giant planets discovered from RV \citep{fischer_planetary_2001, fischer_planetary_2003}. The inner planet (b, $P\sim14$ d, $\msini\sim0.8~\Mj$) can be classified as a `warm Jupiter,' a class of gas giant planets that are thought to orbit too far from their host stars to experience efficient tidal migration and become hot Jupiters. The outer planet (c, $P\sim2136$ d, $\msini\sim13~\Mj$) has been confirmed with multiple independent surveys, and was also studied with HST astrometry by \citet{benedict_mass_2010} who constrained the full orbit of the outer planet. In \S\ref{sec:orbit_results}, we compare our measurements of the outer planet's orbit with those from \citet{benedict_mass_2010}.

\subsection{The debris disc}
For HD~38529, infrared excess emission indicative of a debris disc has been detected at 70$\mu$m ($4.7\sigma$) and marginally at 33$\mu$m ($2.6\sigma$) by {\it Spitzer} \citep{moro-martin_are_2007}. Based on fits to the spectral-energy distribution and dynamical analysis of the planetesimals in the disc, \citet{moro-martin_dust_2007} find that the planetesimals can be stably located between $0.4-0.8$, $20-50$, or beyond 60 au. The star has also been observed by {\it Herschel}, and is resolved at 70$\mu$m by PACS (\texttt{OT1\_gbryden\_1}) with a high significance of $\chi\sim21$ \citep{yelverton_no_2020}. In \S\ref{sec:disc_model}, we model the resolved images of the debris disc in HD~38529 to measure its orientation and size.

\section{Debris disc geometry}\label{sec:disc_model}
In this section, we model the {\it Herschel} PACS 70$\mu$m images of HD~113337 and HD~38529 to constrain the disc geometry for both systems. In particular, we are interested in the inclination, position angle, inner disc radius, and outer disc radius of the debris discs. We briefly overview the method below. For more details, see the appendix of \citet{yelverton_statistically_2019}.

\subsection{Overview of method}
We model the debris disc images using the method described in \citet{yelverton_statistically_2019}.\footnote{The modelling code is freely available at \url{https://github.com/bmy21/pacs-model}.} The model assumes a fixed $r^{-1.5}$ power-law surface brightness profile between inner ($r_{\rm in}$) and outer ($r_{\rm out}$) edges, as well as an inclination $I$ and position angle PA. In addition, we model the resolved discs with a flux of $F_{\rm disc}$, and include two sky offset parameters ($x_0$ and $y_0$) to define the star's location in the image, given the uncertainty in Herschel’s pointing. The discs are also assumed to have negligible scale heights, given the lack of constraints on this from the low-resolution of the images. In total, there are seven parameters for the disc model. The models are created at high resolution, then rebinned to the PACS resolution and convolved with a point-spread function (an observation of a bright calibration star), and subtracted from the data to compute a $\chi^2$ value in a $40\arcsec \times 40$\arcsec region cropped from the full image. To compute the $\chi^2$, we use the pixel rms in nearby pixels multiplied by a factor 2.4 to account for the correlation between pixels \citep{fruchter_drizzle_2002, kennedy_99_2012}. To fit models we use the Markov Chain Monte-Carlo method implemented by the \texttt{emcee} package \citep{foreman-mackey_emcee:_2013}, running 250 parallel chains (“walkers”) for 1000 steps, and use the last 100 steps as posterior distributions for our parameters. All parameters have uniform priors, though $r_{\rm out}$ is restricted to be greater than $r_{\rm in}$.

\begin{table}
\centering
\caption{Results of disc modelling for HD~113337 and HD~38529, after combining posteriors from four different PSFs. We list the median and $1\sigma$ intervals for the resolved disc flux ($F_{\rm{disc}}$), inner disc radius ($r_{\rm{in}}$), outer disc radius ($r_{\rm{out}}$), inclination ($I$), and position angle (PA).}
\label{tab:disc_results}
\begin{tabular}{lccccc}
\hline
\vspace{5pt}
Name & $F_{\rm{disc}}$ (mJy) & $r_{\rm{in}}$ (au) & $r_{\rm{out}}$ (au) & $I$ ($\degr$) & PA ($\degr$) \\
HD~113337 & $170\pm2$ & $19^{+10}_{-8}$ & $174\pm19$ & $13^{+10}_{-9}$ & $-20^{+63}_{-38}$ \\
HD~38529 & $70\pm2$ & $46^{+38}_{-27}$ & $208\pm54$ & $71^{+10}_{-7}$ & $48\pm5$ \\
\hline
\end{tabular}
\end{table}

\begin{table*}
\setlength{\tabcolsep}{8pt}
\centering
\caption{Proper motion anomalies in declination and right ascension for HD~113337 and HD~38529 from \citet{brandt_erratum_2019}. The $\Delta \mu_\alpha$ components have the $\cos{\delta}$ factor included. $\sigma[\Delta\mu]$ represent the uncertainties. The last three columns give the amplitude of the PMa, the uncertainty on the amplitude, and the signal-to-noise ratio. We assume that the uncertainties on the proper motions and the mean motion vector are independent and add them in quadrature to calculate these uncertainties.}
\label{tab:pma_data}
\begin{tabular}{lcccccccc}
\hline
Name & Data & $\Delta\mu_{\delta}$ & $\sigma[\Delta\mu_{\delta}]$ & $\Delta\mu_{\alpha}$ & $\sigma[\Delta\mu_{\alpha}]$ & $\Delta\mu$ & $\sigma[\Delta\mu]$ & S/N\\
& epoch & \multicolumn{2}{c}{\masyr} & \multicolumn{2}{c}{\masyr} & \multicolumn{2}{c}{\ms}\\
\hline
HD~113337 & Gaia & 0.450 & 0.111 & -0.650 & 0.118 & 135.7 & 19.9 & 6.8 \\
HD~113337 & Hip & 0.695 & 0.385 & 1.238 & 0.444 & 243.7 & 74.0 & 3.3 \\
HD~38529 & Gaia & -0.658 & 0.172 & 0.250 & 0.174 & 141.5 & 34.6 & 4.1 \\
HD~38529 & Hip & -0.259 & 0.536 & -1.428 & 0.661 & 291.5 & 132.2 & 2.2 \\
\hline
\end{tabular}
\end{table*}

The one difference relative to \citep{yelverton_statistically_2019} is that we allow for PSF variation, by repeating the fitting four times with different calibration observations. As shown by \citet{kennedy_coplanar_2012}, the PACS PSF varies but the underlying reason is unclear, so this introduces additional uncertainty to the model parameters - here we simply combine the results of all four fitting runs to produce our final posterior distributions, so are essentially assuming that each of the four PSFs used is equally likely to apply at the time of the science observation. The discs here are moderately well resolved, so the inclusion of multiple PSFs does not inflate our uncertainties significantly. To illustrate our fits, we plot the {\it Herschel} PACS 70$\mu$m image, PSF subtracted image, disc model, and residuals in Fig.~\ref{fig:disc_images} for HD~113337 (top panel) and HD~38529 (bottom panel), using one of the four PSFs for the subtraction.

\subsection{Results}
For the HD~113337 disc, we find $I=13^{+10}_{-9}\degr$, PA=$-20^{+63}_{-38}\degr$. The PA is not well constrained because the disc is close to face-on. We find the disc is located between $r_{\rm{in}}=19^{+10}_{-8}$ au and $r_{\rm{out}}=174\pm19$ au. In comparison, \citet{borgniet_constraints_2019} modelled the same {\it Herschel} images and found $I=25\degr^{+5\degr}_{-15\degr}$, PA=$128\pm5\degr$. Although the $I$ measurements are consistent (and the PA could be consistent given the $180\degr$ degeneracy in disc PA values), \citet{borgniet_constraints_2019} quotes much smaller errors for the disc PA. It is possible that their fit did not take into account the covariance between the parameters, the possible impact of PSF variation, and/or correlated noise, as our MCMC fit does, explaining their significantly smaller error bars. The HD~113337 disc is close to face-on, so the disc PA should be relatively unconstrained. For these reasons, and to be consistent with our disc modelling for HD~38529, we choose to use our disc results for the analysis in this paper.

We also fit the debris disc in HD~38529 based on {\it Herschel} images, and find $I=71^{+10}_{-7}\degr$, PA=$48\pm5\degr$. The PA is much better constrained compared to the HD~113337 disc, due to the fact that the geometry is closer to edge-on. We find that the disc is located between $r_{\rm{in}}=46^{+38}_{-27}$ au and $r_{\rm{out}}=208\pm54$ au. The disc location is consistent with the outer stable region ($>60$ au) from \citet{moro-martin_dust_2007}.

\section{Orbits of the outer planets}\label{sec:companion_orbits}

\subsection{The proper motion anomaly}\label{sec:pma_fit}
By combining absolute astrometry from {\it Gaia} DR2 and {\it Hipparcos} with a RV time series, we constrain the full three-dimensional (3-D) orbit of the outer planets in HD~113337 and HD~38529. The astrometric data we use can be called the ``proper motion anomaly'' \citep{brandt_hipparcos-gaia_2018, kervella_stellar_2019}. In short, proper motion anomalies (PMa) are computed by comparing the proper motions from {\it Gaia} DR2 and {\it Hipparcos} with the mean motion vector between the two epochs, as determined by their positional differences \citep{brandt_hipparcos-gaia_2018, kervella_stellar_2019}. In this paper, we use the data as compiled by \citet{brandt_hipparcos-gaia_2018},\footnote{Specifically, the corrected data published in the erratum \citep{brandt_erratum_2019}.} which uses a composite {\it Hipparcos} catalogue and also places {\it Hipparcos} astrometry into the reference frame of {\it Gaia} DR2. Brown dwarf companions with independent orbital measurements have been used to validate the PMa method \citep{brandt_precise_2019, xuan_evidence_2020}. In the planetary regime, the method has been successfully applied on $\epsilon$ Indi Ab \citep{feng_detection_2019}, as well as $\pi$ Men b \citep{xuan_evidence_2020, damasso_precise_2020-1, de_rosa_significant_2020-1} and HAT-P-11 c \citep{xuan_evidence_2020}. Data for the PMa of our targets are given in Table~\ref{tab:pma_data}. 

To model the PMa data, we follow the procedure in \citet{xuan_evidence_2020}, which takes into account the fact that the PMa measurements are not instantaneous velocities, but need to be corrected by using the individual observation times of each mission. When fitting the PMa, we can ignore the contribution of the inner planets and the stellar companions for both targets, as justified in the subsections \S\ref{sec:ignore_inners} and \S\ref{sec:ignore_binary} below.

As shown in Table~\ref{tab:pma_data}, the PMa amplitude of HD~113337 is $\sim137~\ms$ ($6.8\sigma$) at the {\it Gaia} DR2 epoch, and $\sim244~\ms$ ($3.3\sigma$) at the {\it Hipparcos} epoch, comparable to the RV semi-amplitude of $\sim80~\ms$ for HD~113337 c, although larger, suggesting a more face-on orbit. Assuming the lack of additional long-period planets in the system, which is ruled out to some extent by the lack of long-term RV trends and non-detection from imaging \citep{borgniet_constraints_2019}, the PMa is well-explained by perturbations from HD~113337 c. For HD~38529, the PMa amplitude is $\sim142~\ms$ at the {\it Gaia} DR2 epoch ($4.1\sigma$), and $\sim292~\ms$ ($2.2\sigma$) at the {\it Hipparcos} epoch, again comparable to the RV semi-amplitude of $\sim172~\ms$ for HD~38529 c. The system has been monitored in RV for over 15 yr without evidence for additional planets, so we can safely assume that the PMa arises from HD~38529~c alone.

\begin{figure*}
    \centering
\begin{subfigure}
    \centering
    \includegraphics[width=.4\linewidth]{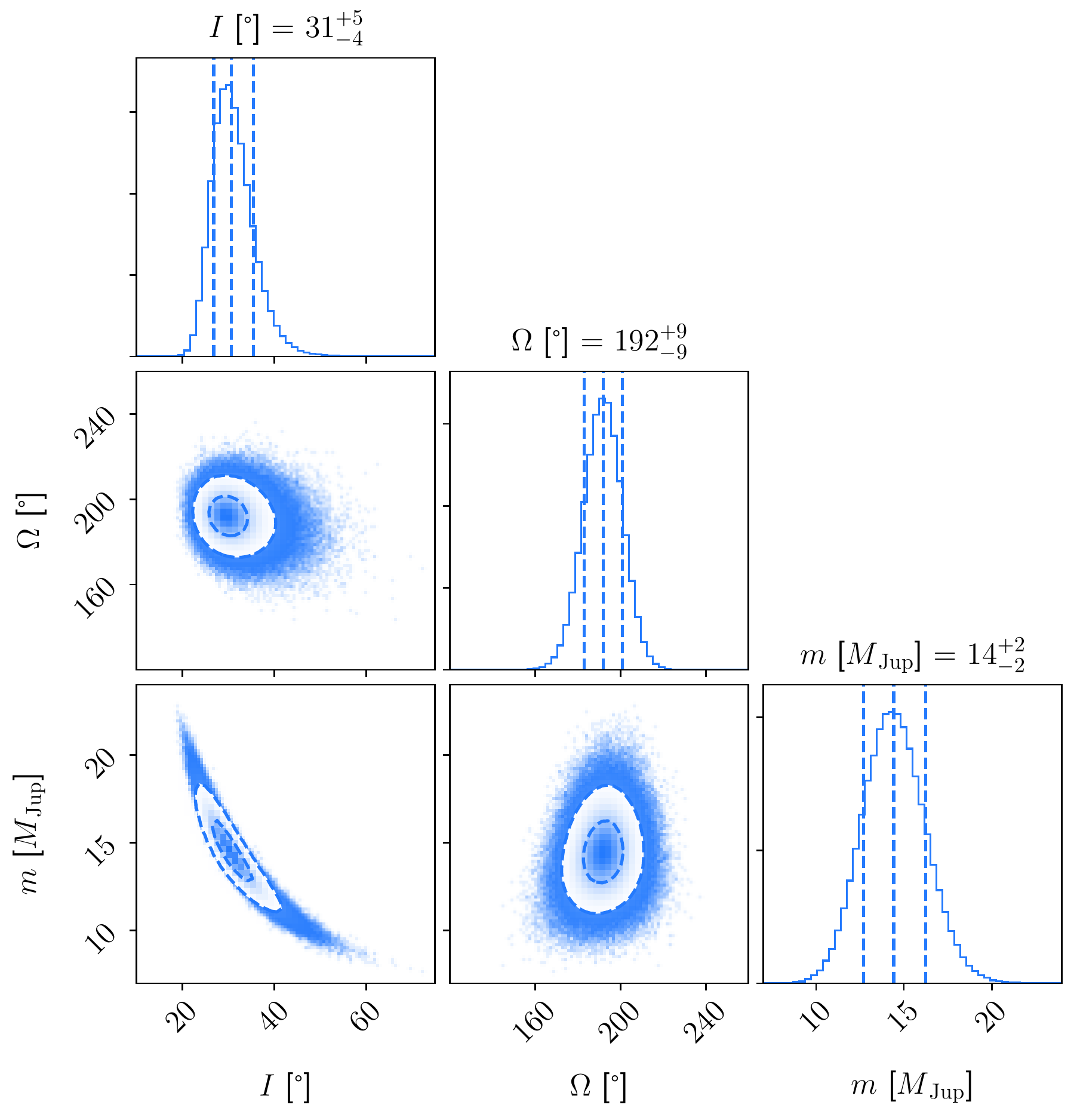}
\end{subfigure}
\begin{subfigure}
    \centering
    \includegraphics[width=.4\linewidth]{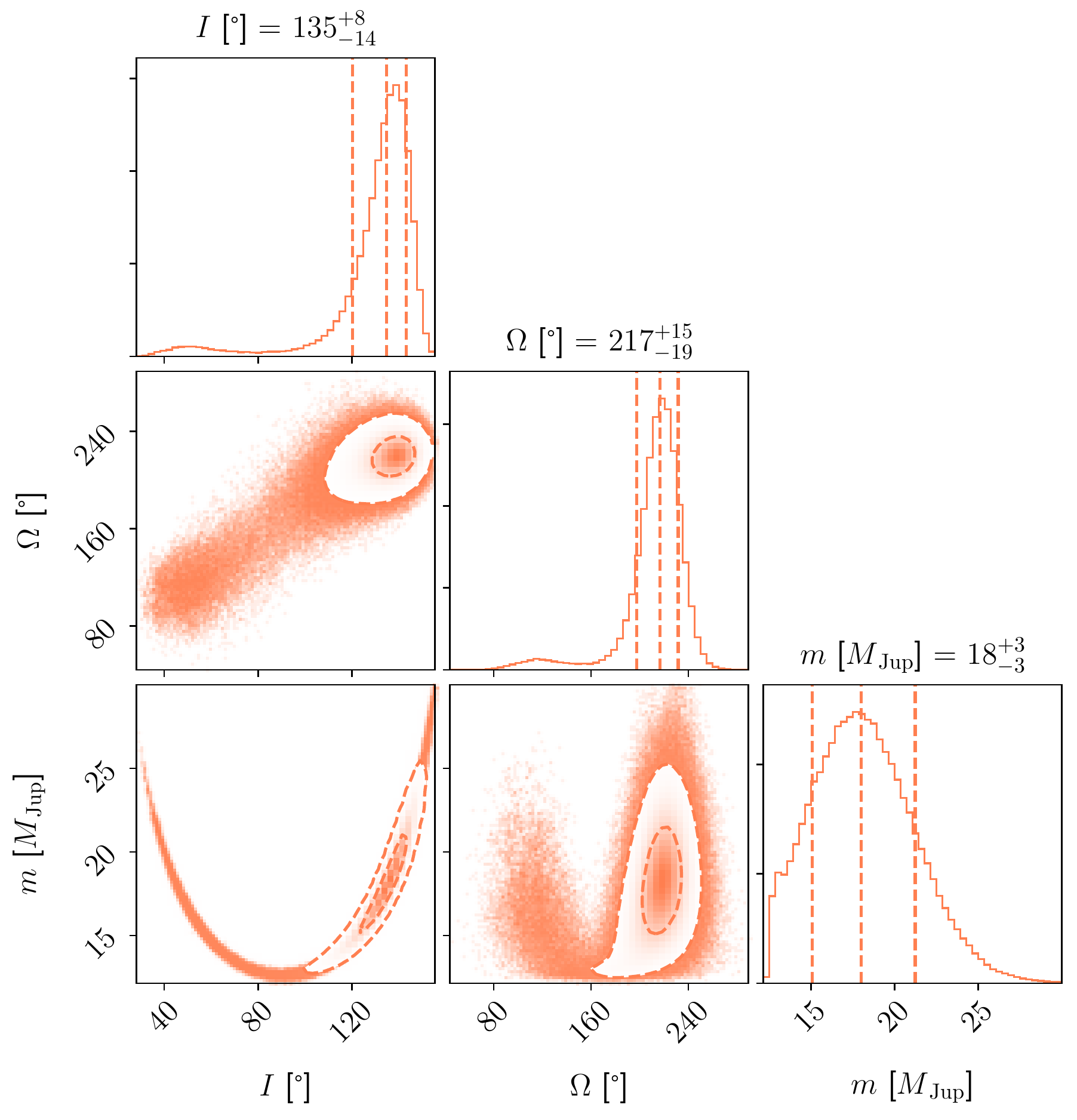}
\end{subfigure}
    \caption{Joint posterior distributions of $I$, $\Omega$, and $m$ for HD~113337~c (left) and HD~38529  (right). Moving outward, the dashed lines on the 2D histograms correspond to $1\sigma$ and $2\sigma$ contours.}
    \label{fig:corner_3param}
\end{figure*}

\subsection{Negligible contribution from inner planets}\label{sec:ignore_inners}
The inner planets in both systems can be ignored when modelling the PMa because of their short periods. Specifically, objects with orbital periods much shorter than the data collection time spans of {\it Gaia} DR2 and {\it Hipparcos} will have their signal smoothed over many orbits during the two observation windows, and therefore averaged to negligible amounts \citep{kervella_stellar_2019}.

The orbital period of HD~38529~b is $\sim14$ d, more than an order of magnitude shorter than the data collection time spans of {\it Gaia} DR2 and {\it Hipparcos} for this target, which are $\sim551$ d and  $\sim 1102$ d, respectively. Therefore, HD~38529~b has a negligible influence on the observed PMa for HD~38529. The period of HD~113337~b is close to 1 yr ($P\sim323$ d), comparable to the data collection times. However, this period is close to 1/2 of the {\it Gaia} DR2 collection span for HD~113337 (628 d). This means that the PMa signal from the inner planet will be smoothed over approximately 2 orbital periods during the {\it Gaia} DR2 observation window, and therefore be averaged out to nearly zero. The {\it Hipparcos} data was collected over about 1125 d for HD 113377, or about 3.5 orbital periods. In this case, orbital smearing will typically reduce the signal to $<0.1$ of its original amplitude, as demonstrated statistically in \citet{kervella_stellar_2019} (see their fig. 2). The PMa amplitude, which scales as $m a^{-1/2}$, is very similar for the outer and inner planets in HD~113337. This means that the contribution from HD~113337~b can be ignored at the 10 per cent level. This is smaller than the size of the uncertainty in the HD~113337 PMa, which has an average uncertainty of 23 per cent (see Table~\ref{tab:pma_data}).

\begin{table*}
\footnotesize
\centering
\setlength{\tabcolsep}{4pt}
\caption{Orbital parameters for HD~113337~c and HD~38529~c from our fits. We also list the adopted stellar masses and {\it Gaia} DR2 parallaxes.}
\label{tab:pma_fit}
\begin{tabular}{lcccccccccc}
Name & $I$ ($\degr$) & $\Omega$ ($\degr$) & $m$ ($\Mj$) & $P$ (d) & $e$ & $\omega_\star$ ($\degr$) & $T_p$ (BJD) & $M_\star (\Msun)$ & $\pi$ (mas) \\
\hline\noalign{\vskip 2pt}
HD~113337 c & $31^{+5}_{-4}$ & $192\pm9$ & $14\pm2$ & $3165^{+70}_{-59}$ & $0.06^{+0.03}_{-0.04}$ & $161^{+28}_{-23}$ & $2458443^{+242}_{-263}$ & $1.40\pm0.04^a$ & $27.640\pm0.051$ \\
HD~38529 c & $135^{+8}_{-14}$ & $217^{+15}_{-19}$ & $18\pm3$ & $2135\pm2$ & $0.34\pm0.01$ & $-161\pm1$ & $2452259\pm7$ & $1.36\pm0.02$ & $23.611\pm0.067$ \\
\hline
\multicolumn{11}{p{0.9\linewidth}}{$^a$ For HD~113337, we adopt the stellar mass that corresponds to the old age solution ($1.40\pm0.04\Msun$). The masses from the two age solutions are consistent within $1\sigma$ (see \S\ref{sec:113337_intro}).}\\
\end{tabular}
\end{table*}

\subsection{Negligible contribution from stellar companions}\label{sec:ignore_binary}
We can also ignore any contribution from the wide M-type stellar companions to the PMa due to their extremely long periods. This is because the reflex motion due to a wide-orbiting companion is essentially linear and constant on the 24.25 yr time-scale between {\it Hipparcos} and {\it Gaia} DR2. Any linear and constant motion is subtracted away when computing the PMa, as it would be absorbed into the mean motion vector between the two epochs. This is demonstrated statistically in fig. 3 of \citet{kervella_stellar_2019}.

For HD~113337, the projected separation of the binary pair of M dwarfs is $\sim4000$ au, which would translate to a period of more than \SI{1.8e5} yr, assuming a semimajor axis of $4000$ au, a primary mass of 1.4$\Msun$ \citep{borgniet_constraints_2019} and a total binary mass of 0.5$\Msun$ \citep{janson_astralux_2012}. The 24.25 yr time span between {\it Hipparcos} and {\it Gaia} DR2 is $\sim0.01$ per cent of the orbital period. Therefore, we can ignore contributions from the binary pair of HD~113337 to the primary star's PMa. For HD~38529, the projected separation of the single stellar companion is $\sim12000$ au, so its orbital period would be roughly a factor of five longer. We ignore its contribution to the PMa of HD~38529 for the same reason.

\subsection{Joint fits of PMa and RV data}
By simultaneously fitting the PMa data and a RV time series, we fully constrain orbits of HD~113337 c and HD~38529 c (i.e. the outer planets). We use seven parameters to describe the orbit: the planet's true mass ($m$), orbital period ($P$), time of periastron ($T_p$), inclination ($I$), longitude of ascending node ($\Omega$), eccentricity ($e$), and argument of periastron of the stellar orbit ($\omega_\star$).\footnote{We use the same coordinate system and definitions for the orbital angles as in \citet{xuan_evidence_2020}.} Although the PMa is assumed to arise from the outer planet alone, we include the inner planet when fitting the RV data with five parameters, $P_b$, $T_{p,b}$, $e_b$, $\omega_{b,\star}$, and $K_b$, the RV semi-amplitude of the inner planet. In our fits, $e$ and $\omega$ are fitted as $\sqrt{e}\cos{\omega}$ and $\sqrt{e}\sin{\omega}$, and $I$ is sampled as $\cos{I}$. We include the stellar mass ($M_\star$) and the parallax ($\pi$) in our fits with Gaussian priors, while all other parameters have uniform priors. Lastly, a set of instrumental offset and jitter terms are fitted for each RV instrument.

The log likelihood of our joint RV and PMa fit is given by \citep{brandt_precise_2019, xuan_evidence_2020}
\begin{equation}
    ln\curL = -\frac{1}{2}(\chi^2_{\Delta \mu}+ \chi^2_{RV}).
\label{eq:loglike}
\end{equation}
The PMa part ($\chi^2_{\Delta \mu}$) is given in \citet{xuan_evidence_2020}. The RV log likelihood ($\chi^2_{RV}$) is based on a two-Keplerian model, with additional zero-point offset and jitter terms \citep[see e.g.][]{howard_nasa-uc-uh_2014}.

We run our fits using the Parallel-Tempered Monte Carlo Markov Chain (PTMCMC) implemented in \texttt{emcee} \citep{foreman-mackey_emcee:_2013}, based on the algorithm from \citet{earl_parallel_2005}. We use 40 temperatures for our 16 parameter models, with 50 walkers in each separate MCMC, and run the fits for \SI{6e5} steps including a burn-in of \SI{6e4} steps that are discarded. The results of these fits are described in the next two subsections, and summarized in Table~\ref{tab:pma_fit}.

\begin{figure*}
\centering
\begin{subfigure}
    \centering
    \includegraphics[width=.45\linewidth]{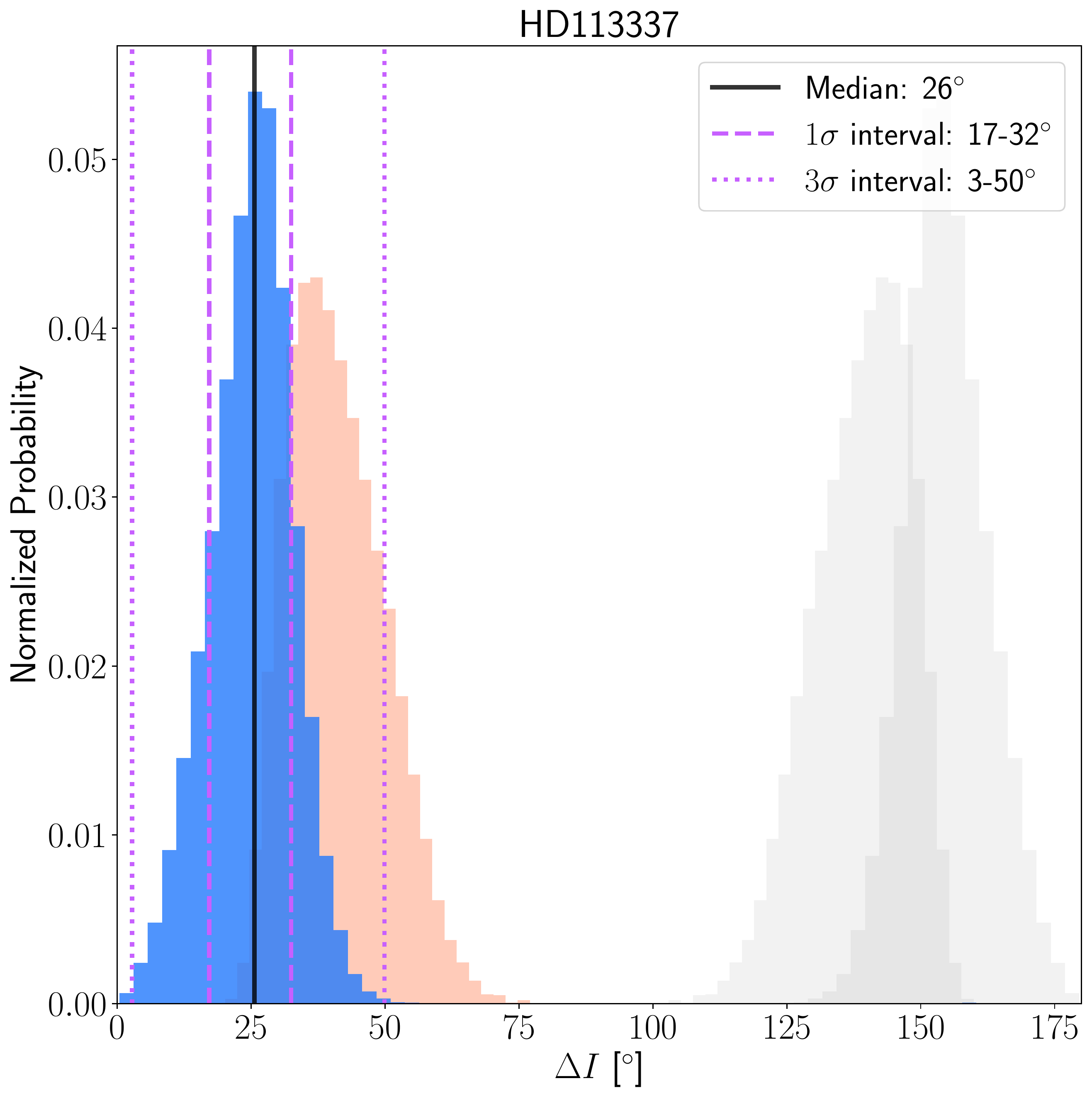}
    \hspace{10mm}
\end{subfigure}
\begin{subfigure}
    \centering
    \includegraphics[width=.45\linewidth]{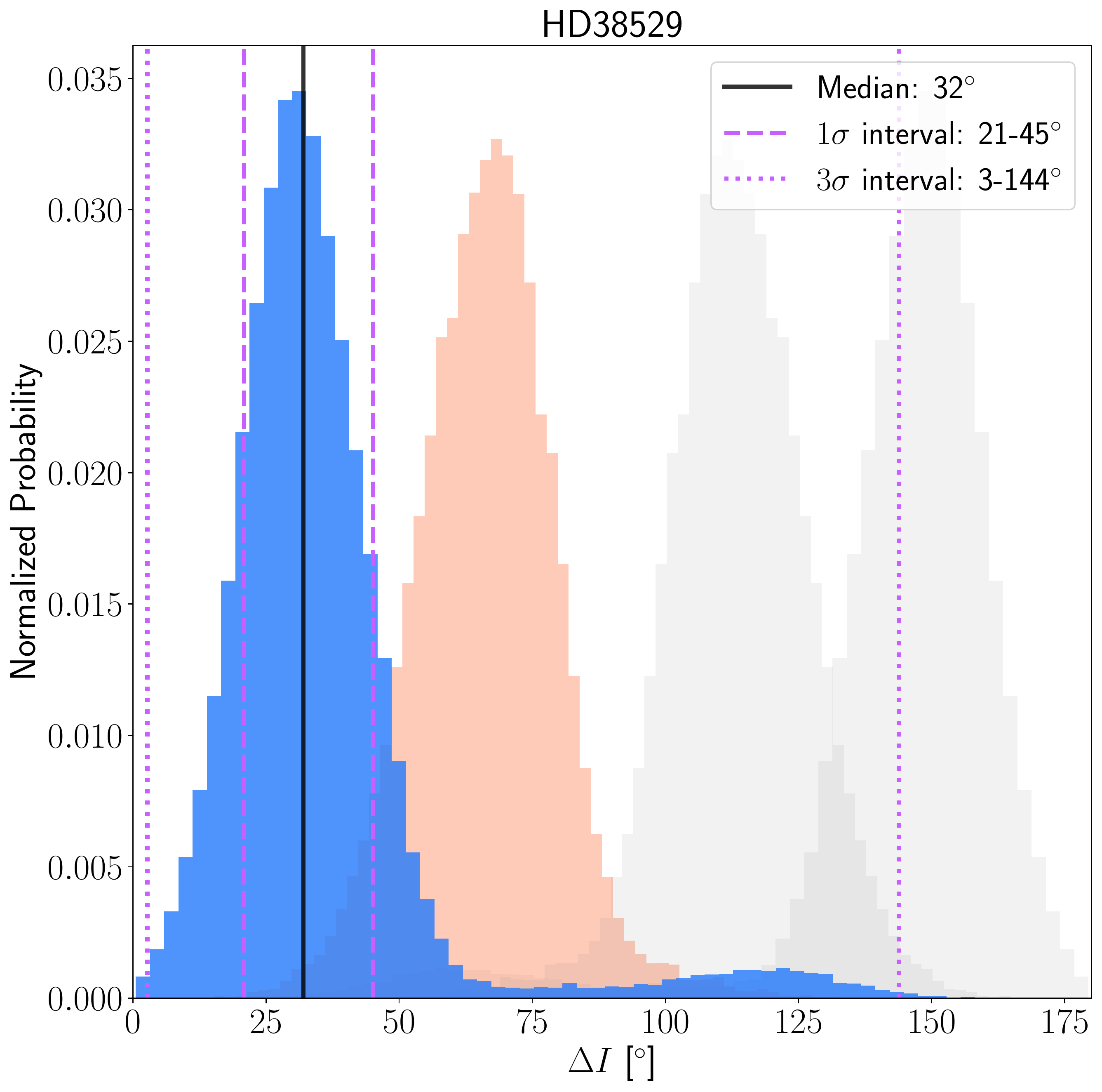}
\end{subfigure}
    \caption{Left: mutual inclination solutions between the orbit of HD~113337~c and its debris disc. The minimum $\Delta I$ solution is shown in blue, with the median value indicated by the black line and $1\sigma$ and $3\sigma$ intervals marked by the dashed and dotted purple lines, respectively. The second prograde solution, corresponding to larger $\Delta I$, is shown in orange. Finally, the two retrograde solutions are plotted in light grey. Right: same for HD~38529~c and its debris disc.}
    \label{fig:Imut}
\end{figure*}

\subsection{Results}\label{sec:orbit_results}
For HD~113337 c, we find $I_c = 31^{+5}_{-4}\degr$, $\Omega_c = 192\pm9\degr$, and $m_c = 14\pm2~\Mj$ (see Fig.~\ref{fig:corner_3param} and Table~\ref{tab:pma_fit}). Measurements for other parameters (including those for the inner planet) are consistent within $1\sigma$ to results from \citet{borgniet_extrasolar_2019},\footnote{There are offsets of $\pi$ between our $\omega_\star$ values and literature values. However, this is merely due to differences in coordinate system definitions (specifically our $Z$ axis points toward the observer), as explained in \citet{xuan_evidence_2020}. The same is true for HD~38529 below.} and we show the full posterior distributions in Appendix~\ref{appendixB}.

For HD~38529 c, we find $I_c = 135^{+8}_{-14}\degr$, $\Omega_c = 217^{+15}_{-19}\degr$, and $m_c = 18\pm3~\Mj$ (see Fig.~\ref{fig:corner_3param} and Table~\ref{tab:pma_fit}). Due to the relatively low S/N of the {\it Hipparcos} epoch PMa data (see Table~\ref{tab:pma_data}), about 5 per cent of the solutions favour a different orbit, as shown by the long tail towards $I_c < 90\degr$ in the right panel of Fig.~\ref{fig:corner_3param}. The full posterior distributions are given in Appendix~\ref{appendixB}, and our measurements for all other parameters are consistent within $1\sigma$ to results from \citet{henry_host_2013}.

We note that \citet{benedict_mass_2010} has measured the orbit of HD~38529~c using the {\it Hubble Space Telescope} (HST) Fine Guidance Sensor, and found $I_c = 49\pm4\degr$ and $\Omega_c = 38\pm8\degr$. We find that the orbit shown in fig. 11 of \citet{benedict_mass_2010} is inconsistent with the RV time series, and likely has orbital angles off by $\pi$ (F. Benedict, private communication). If their orbital direction is reversed from counterclockwise (as seen by the observer) to clockwise, this corrects the offsets. After correction, their measurement would be $I_c = 131\pm4\degr$ and $\Omega_c = 218\pm8\degr$, consistent with our values to $<1\sigma$. Due to possible confusion in the HST measurement, we use our results from the PMa and RV joint fit, despite the higher uncertainties in our measurements.

From our full orbit fits, we find that both HD~113337~c and HD~38529~c have true masses that exceed the standard deuterium-burning mass of $13~\Mj$, which is the traditional dividing line between planets and brown dwarfs. However, the definitions of planets and brown dwarfs are undergoing debate \citep[e.g.][]{chabrier_giant_2014, schlaufman_evidence_2018}. Due to this uncertainty, we will continue to call them planets in this paper.

\section{Assessing system alignment}\label{sec:assess_align}
In this section, we first compute the mutual inclinations between the outer planets and debris discs in HD~113337 and HD~38529 (\S\ref{sec:deltaI}) based on results from \S\ref{sec:disc_model} and \S\ref{sec:companion_orbits}. Then, in \S\ref{sec:sec_evol}, we interpret the mutual inclinations by calculating the forced inclination of planetesimals in the disc, and the associated secular precession time-scale of the disc. Finally, we examine whether the stellar spin axis is aligned with either the outer planets or the debris discs. We note that unless otherwise specified, we refer to the outer planets when using the word `planet' in this section.

\subsection{Planet-disc mutual inclinations}\label{sec:deltaI}
The mutual inclination between the outer planet's orbit and the observed disc mid-plane is given by
\begin{equation}\label{eq:deltaI}
    \cos{\Delta I} = \cos{I_c}\cos{I_{\rm{disc}}} + \sin{I_c}\sin{I_{\rm{disc}}}\cos{(\Omega_c - \Omega_{\rm{disc}})},
\end{equation}
where $I_c$ and $\Omega_c$ are inclination and longitude of ascending node of the outer planets (planet c in both systems), and $I_{\rm{disc}}$ and $\Omega_{\rm{disc}}$ represent the same quantities for the disc.

We note that analysis of resolved disc images leaves two uncertain aspects of the disc orientation. First, the direction that material orbits in the disc is a priori unknown, which gives rise to a degeneracy between $I_{\rm{disc}}$ and $\pi-I_{\rm{disc}}$ (i.e. whether the disc angular momentum points toward or away from the observer). Second, which side of the disc is above the sky plane and which side is below is unknown, which amounts to a degeneracy between the $\Omega_{\rm{disc}}$ = PA and $\Omega_{\rm{disc}}$ = PA + $\pi$. Without additional constraints on the disc, there are then $2\times2=4$ possible directions for the disc angular momentum vector to point, and consequently four possible values of the planet-disc $\Delta I$ (even if the planet's orbit is uniquely constrained). 

For all four cases, we randomly sample from distributions of the parameters in Eq.~\ref{eq:deltaI} to construct four distinct distributions of $\Delta I$. For $\Omega_c$ and $I_c$, we draw from the posterior distributions of the orbit fits, while for $\Omega_{\rm{disc}}$ and $I_{\rm{disc}}$, we draw from posteriors of the disc modelling, taking into account the adjustments described above to yield four solutions of $\Delta I$. The sampling process yields two prograde solutions and two retrograde solutions that are mirror reflections of the prograde solutions about $\Delta I=90\degr$.

We plot the $\Delta I$ posteriors in Fig.~\ref{fig:Imut} for HD~113337 (left panel) and HD~38529 (right panel). For both figures, the two prograde cases are shown in blue and orange, where the blue histogram corresponds to the minimum or most aligned $\Delta I$ solution. The median as well as $1\sigma$ and $3\sigma$ confidence intervals are overlaid for the most aligned $\Delta I$ solution, which is the only solution that could be consistent with planet-disc alignment (with $<0.3$ per cent chance). The two retrograde solutions are shown in light grey. It might be physically unlikely for debris discs to be orbiting retrograde relative to the orbits of the planets, so we focus on the prograde solutions, and especially the most aligned $\Delta I$ solution which permits richer discussion on both nearly aligned and misaligned possibilities.

For HD~113337, we find that the most aligned planet-disc $\Delta I=17-32\degr$ at $1\sigma$, $\Delta I =9-40\degr$ at $2\sigma$, and $3-50\degr$ at $3\sigma$. Therefore, there is evidence for misalignment, which is due to slight offsets in both the $I$ and $\Omega$ distributions between the two orbits. Specifically, $I_c=31^{+5}_{-4}\degr$ and $I_{\rm{disc}}=13^{+10}_{-9}\degr$, while $\Omega_c = 192\pm9\degr$ and $\Omega_{\rm{disc}}=160^{+63}_{-38}\degr$ (see \S\ref{sec:disc_model}, \S\ref{sec:orbit_results}), where the quoted disc orientation is that which gives the minimum $\Delta I$ solution. The other $\Delta I$ solutions correspond to larger misalignments. For example, the second prograde solution has $\Delta I=32-51\degr$ at $1\sigma$.

For HD~38529, the most aligned planet-disc $\Delta I=21-45\degr$ at $1\sigma$, $\Delta I=10-116\degr$ at $2\sigma$, and $3-144\degr$ at $3\sigma$. The long tail to the right is caused by the small cluster of planet solutions that favour a different orbit (see Fig.~\ref{fig:corner_3param}). The misalignment is again driven by slight differences in both $I$ and $\Omega$. Specifically, $I_c=135^{+8}_{-14}\degr$ and $I_{\rm{disc}}=109^{+10}_{-7}\degr$, while $\Omega_c=217^{+15}_{-19}\degr$ and $\Omega_{\rm{disc}}=228\pm5\degr$ (\S\ref{sec:disc_model}, \S\ref{sec:orbit_results}). For comparison, the second prograde solution has $\Delta I=54-78\degr$ at $1\sigma$. 

Given crude constraints on the orbit of HD~38529~B in \S\ref{sec:38529_intro}, we calculate the mutual inclination between HD~38529~B and HD~38529~c and find that the planet-binary $\Delta I$ is larger than $20\degr$ at the $3\sigma$ level. While we do not draw quantitative conclusions from this due to the strong prior-dependence of the binary orbital parameters (see \S\ref{sec:38529_intro} for details), a large misalignment between HD~38529~c and B is plausible.

If the disc orientations are such that the most aligned $\Delta I$ solutions are the real ones, then the distributions for HD~38529 and HD~113337 are similar, and both systems are inconsistent with alignment at $\gtrsim3\sigma$ confidence according to our measurements. The second prograde $\Delta I$ solutions imply even larger misalignments, while the two other solutions would imply that the planet and disc orbit in retrograde.

\subsection{Secular evolution of the disc}\label{sec:sec_evol}

\subsubsection{Basic motion and forced inclination}

\begin{figure}
    \centering
    \includegraphics[width=1.0\linewidth]{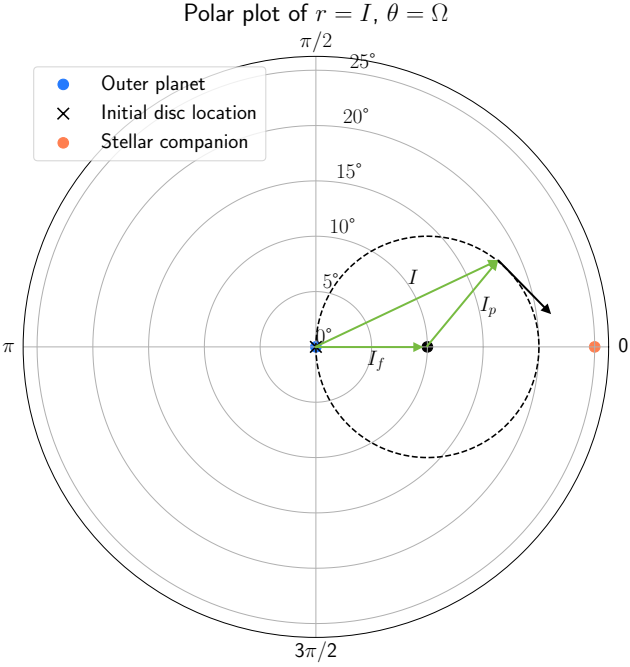}
    \caption{Illustration of precession of disc particles at the same semimajor axis about the forced inclination vector, defined by $I_f$. Each location on this polar plot of $r=I$, $\theta=\Omega$ defines an orbital plane. Here, $I$ and $\Omega$ are measured relative to the planet's orbit which is at the origin (blue dot). The orbit of the stellar companion is set at $I=25\degr$, $\Omega=0$ (orange dot). Orbits of disc particles are assumed to start aligned with the planet's orbit (i.e. black cross at origin); then, secular precession causes their orbits to precess clockwise around a plane defined by $I_f$ (black dot) that lies in between orbits of the planet and stellar companion. The green arrows denote the disc particles' forced and proper inclination vectors (with magnitudes $I_f$ and $I_p$) as well as instantaneous inclination (with magnitude $I$) at one specific instance of the precession cycle. Because the particles start at the origin with $I=0$, they precess with $I_p=I_f$ around the forced inclination vector.}
    \label{fig:complexI}
\end{figure}

\begin{figure*}
\centering
\begin{subfigure}
    \centering
    \includegraphics[width=0.48\linewidth]{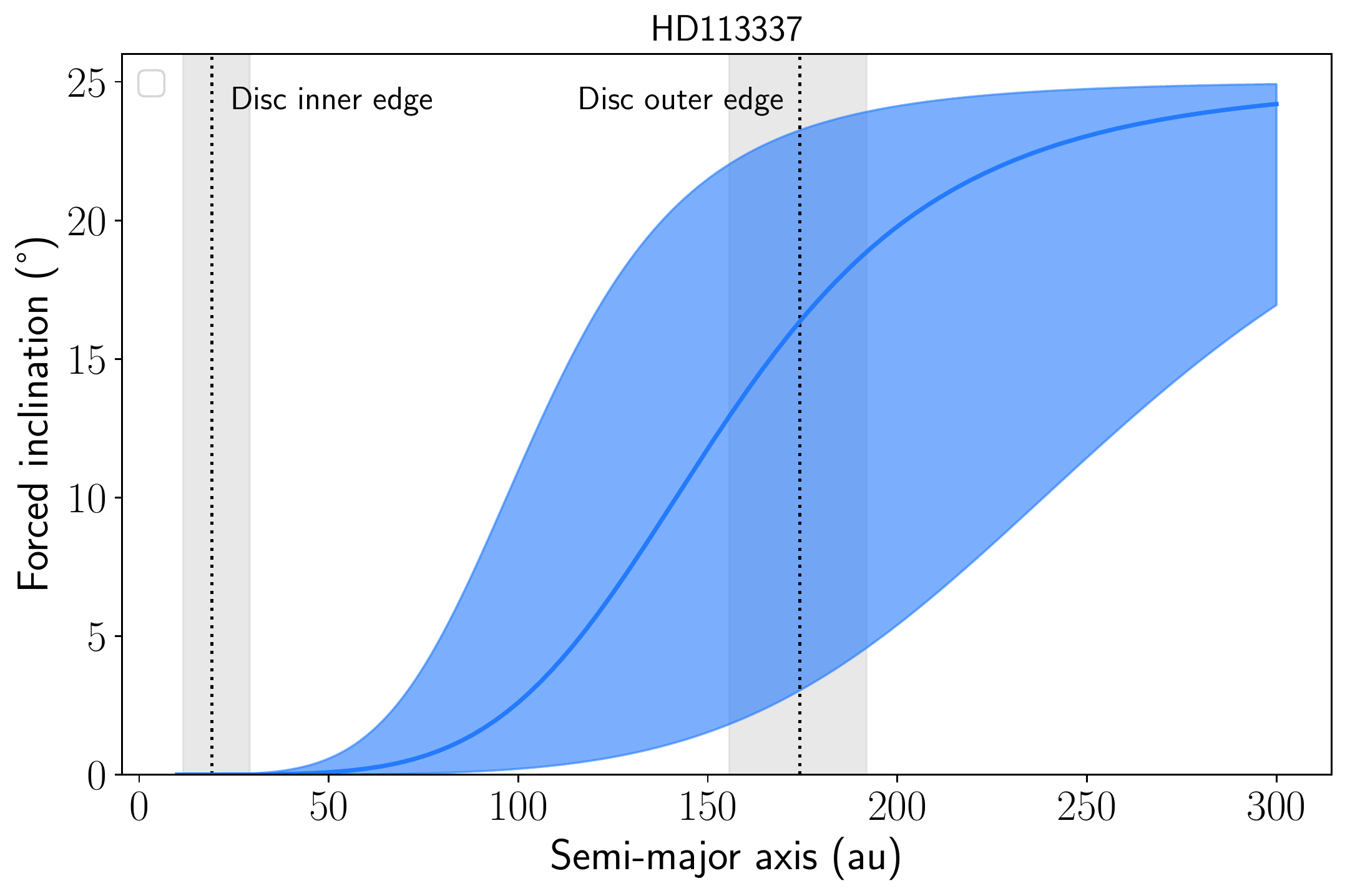}
    \end{subfigure}
\begin{subfigure}
    \centering
    \includegraphics[width=0.48\linewidth]{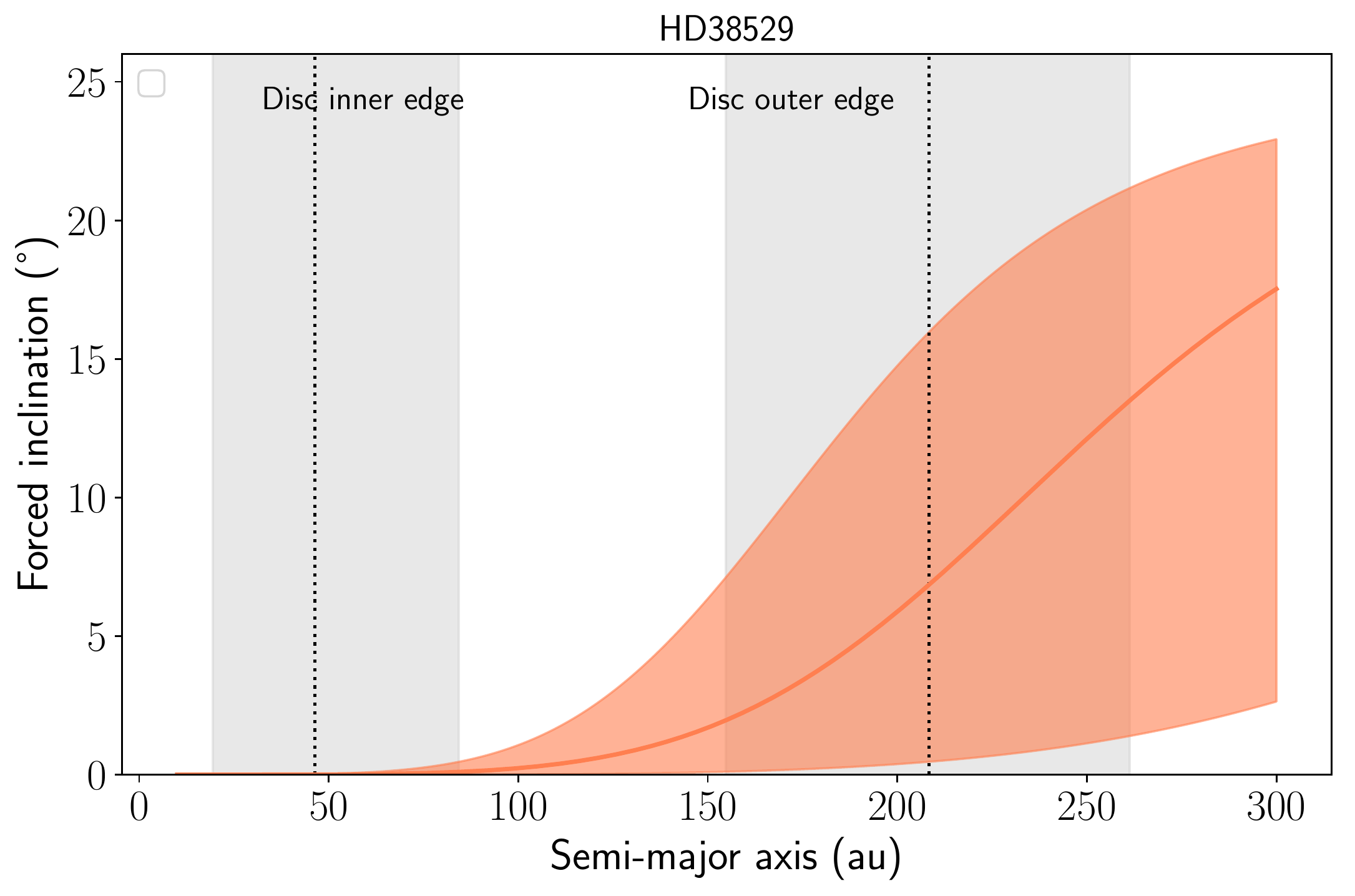}
\end{subfigure}
    \caption{Forced inclination of planetesimals in the debris disc under secular perturbations by the outer planets and stellar companions for HD~113337 (left) and HD~38529 (right). The solid lines trace the median $I_f$ and the shaded regions indicate the $2\sigma$ interval, which are generated by considering uncertainties in $a$ and $m$ of the planets and binaries, but fixing the planet-binary $\Delta I$ to be $25\degr$. In this figure, the planet has $I=0$, and the binary has $I=25\degr$. The $1\sigma$ intervals for the disc inner ($r_{\rm in}$) and outer ($r_{\rm out}$) edges are indicated by the grey regions, and the median $r_{\rm in}$ and $r_{\rm out}$ values are plotted as black dotted lines.}
    \label{fig:forcedI}
\end{figure*}

In this section, we examine how the planets and stellar companions of HD 113337 and HD 38529 influence their debris discs, and whether the discs are expected to be aligned with the outer planets or not. Specifically, we consider the evolution of planetesimals in the disc by treating them as test particles that evolve due to secular perturbations from the planets and stellar companions. We use the Laplace-Lagrange theory, which is valid in the regime where inclinations and eccentricities are small, and is appropriate for our minimum planet-disc $\Delta I$ solutions from \S\ref{sec:deltaI}. The stellar companions have unconstrained orbits, so the discussion below also assumes that they have small inclinations and eccentricities.

Under the Laplace-Lagrange theory, the semimajor axis ($a$) of a massless disc particle remains constant, while variations in its eccentricities and inclinations ($e$ and $I$) are coupled with variations in the longitude of pericentre and ascending node ($\varpi$ and $\Omega$). Here, we focus on the evolution of $I$ and $\Omega$, which is best described with the complex inclination $I \exp^{i \Omega}$ (where $i$ is the unit imaginary number). The secular evolution of the complex inclination can be decomposed into `forced' and `proper' components, which are added as vectors in the complex plane (see Fig.~\ref{fig:forcedI}). The forced component is characterized by an amplitude called the forced inclination ($I_f$), which depends on the perturber orbits and the semimajor axis of the particle. On the other hand, the proper inclination ($I_p$) is the magnitude of the constant-length vector between the forced inclination and instantaneous inclination vectors. Due to their short orbital distances and lower masses, we find that the inner planets (HD~113337~b, HD~38529~b) induce precession time-scales at the location of the disc that that are orders of magnitude longer than that of the outer planets (HD~113337~c, HD~38529~c), so we ignore them in this analysis. For each system, there are then two perturbers, which impose a forced inclination on the disc particles given by \citep[e.g.][]{wyatt_how_1999}
\begin{equation}\label{eq:If}
    I_f = \frac{B_2I_2-B_1I_1}{B_{12}+B_{21}-B_1-B_2},
\end{equation}
where we use the subscripts $1$ and $2$ to denote the outer planet and stellar companion, respectively. For example, $I_1$ is the inclination of the outer planet. $B_1$ and $B_2$ are coefficients in the disturbing function that set the particle's precession rate due to each perturber, and can be generalized as $B_j$ \citep{murray_solar_2000}
\begin{equation}\label{eq:Bj}
    B_j = \frac{n}{4} \frac{m_j}{M_\star}\alpha_j\bar{\alpha}_j b^{(1)}_{3/2}(\alpha_j),
\end{equation}
where $n=2\pi/P$ is the mean motion of the particle, $m_j$ and $M_\star$ are the masses of the perturber and central star, and $\alpha_j$ and $\bar{\alpha}_j$ are defined such as for an interior perturber (the planets in this case), $\alpha_j=a_j/a$ and $\bar{\alpha}_j=1$, whereas for an exterior perturber (the stellar companions), $\alpha_j=\bar{\alpha}_j=a/a_j$. $b^{(1)}_{3/2}(\alpha_j)$ is a Laplace coefficient, where $b^{(1)}_{3/2}(\alpha_j)\approx3\alpha_j$ when $\alpha_j \ll 1$. Lastly, the $B_{12}$ and $B_{21}$ coefficients in Eq.~\ref{eq:If} describe the precession rate of each perturber caused by the other \citep{murray_solar_2000}
\begin{equation}\label{eq:Bjk}
    B_{jk} = \frac{n_j}{4} \frac{m_k}{M_\star+m_j}\alpha_{jk}\bar{\alpha}_{jk}b^{(1)}_{3/2}(\alpha_{jk}),
\end{equation}
where the subscripts $j$ and $k$ denote the perturbers, and $\alpha_{jk} = \bar{\alpha}_{jk} = a_j / a_k$ for $a_k > a_j$, while $\alpha_{jk} = a_k / a_j$, $\bar{\alpha}_{jk} = 1$ otherwise. The other parameters are defined as for Eq.~\ref{eq:Bj}. For example, $B_{12}$ is the precession rate of the outer planet caused by the stellar companion. We find that the time-scales associated with $B_{12}$ and $B_{21}$ are orders of magnitude longer than system age for both systems, meaning that the outer planets and stellar companions do not interact significantly via secular perturbations.

We illustrate the precession of disc particles in Fig.~\ref{fig:complexI}, which shows the evolution of the complex inclination in a polar plot of $r=I$, $\theta=\Omega$, where each point on the plot indicates an orbital plane. For simplicity, we set the outer planet at the origin with $I_1=0\degr$ (blue dot). The planet-binary $\Delta I$ is unknown for HD~113337, while for HD~38529 we find the planet-binary $\Delta I$ could be $>20\degr$ at $3\sigma$ (see \S\ref{sec:deltaI}). Overall, it is plausible that the outer planets and binary stars are misaligned, as the binaries are thousands to ten thousands of au away and alignment of components is not expected beyond $\sim100$ au scales \citep{hale_orbital_1994}. As an example, we set the stellar companion at $I_2=25\degr$, $\Omega_2=0$ (orange dot), defined relative to the planet. The initial position of the debris disc is also unknown. Here, we consider the case where the disc is initially aligned with the planet (i.e. black cross at the origin). We also assume the outer planet and stellar companion have fixed positions in this plot, as their mutual perturbations act on time-scales much longer than the system age. As a result, $I_f$ is also fixed. Using Eq.~\ref{eq:If}, we can determine the forced inclination of particles as a function of $a$. As an example, we consider a specific value of $I_f = 10\degr$ (black dot). If we assume the disc particles start at the origin, Laplace-Lagrange theory predicts that their motion in Fig.~\ref{fig:complexI} is to precess clockwise around a circle centred at $I_f$ at a rate equal to $B_1 + B_2$. 

Because the precession rate depends on a particle's $a$ (see Eq.~\ref{eq:Bj}), after a few precession cycles particles with slightly different $a$ will be at random parts of the precession, and therefore be distributed uniformly in the dashed circle in Fig.~\ref{fig:complexI}. Therefore, after a few precession cycles, the mid-plane of the disc at this value of $a$ would have an inclination of $I_f$. Note that this is independent of the initial disc location, as $I_f$ only depends on the perturber orbits and the disc $a$. At the same time, the disc would be vertically puffed by two times the radius of the dashed circle; it would have an angular scale height of $10\degr$ given $I_f=10\degr$ and an initial disc location at the origin. Because $I_f$ is always in the planet and binary orbits, the above scenario could be responsible for misalignments between the outer planets and discs in HD~113337 and HD~38529.

Fig.~\ref{fig:complexI} considers the evolution of disc particles at a given $a$. Using Eq.~\ref{eq:If}, we now calculate $I_f$ between $10-300$ au for HD 113337 and HD 38529. To factor in uncertainties in the perturbers' semimajor axes and masses (the only two parameters used in Eq.~\ref{eq:If}), we calculate $I_f$ for 1000 randomly drawn planet and stellar companion orbits, and derive 1000 values of $a$ and $m$ for each perturber. For the planets, we draw their orbital elements and masses from posterior distributions in \S\ref{sec:pma_fit}. For the stellar companion masses, we impose generous error bars and draw from uniform distributions of $0.5\pm0.1\Msun$ and $0.35\pm0.1\Msun$ for HD~113337~B and HD~38529~B, respectively (see \S\ref{sec:113337_intro}, \S\ref{sec:38529_intro}). We note that HD~113337~B is found to be a binary pair \citep{janson_astralux_2012}, but we treat it as a single star for this estimate and our subsequent discussion on the dynamics. The stellar companion orbits are unknown, so we randomly generate 1000 sets of uniformly distributed orbital elements, drawing the mean anomaly, $\omega$, and $\Omega$ from $0-2\pi$, and $e$ from $0-0.8$ to match the observed eccentricity distribution for stellar binaries \citep{raghavan_survey_2010}. We choose to randomly draw orbits for HD~38529~B as well because its orbital constraints from {\it Gaia} DR2 relative positions and velocities are poor (derived from LOFTI in \S\ref{sec:38529_intro}). For $I$, we draw from -1 to 1 for HD~113337~B, but from -1 to 0 for HD~38529~B, as its orbital direction is constrained (see \S\ref{sec:38529_intro}).\footnote{In accordance with the assumption of a planet-binary $\Delta I=25\degr$, the binary orbits are drawn so that the binary's $I$ is always $25\degr$ larger than the planet's, while the binary's $\Omega$ is the same as the planet's.} With each set of binary orbital elements, we compute a simulated ratio of its true $a$ to the expected projected separation, and multiply this ratio by the observed projected separation to get a value of $a$ for the binary.

We plot the median $I_f$ and associated $2\sigma$ regions (i.e. 95 per cent confidence intervals) in Fig.~\ref{fig:forcedI} for HD~113337 (left panel) and HD~38529 (right panel). Following the assumption used for Fig.~\ref{fig:complexI}, we set $I_1=0\degr$ for the outer planet and $I_2=25\degr$ for the stellar companion. To show the disc extent, we plot the median $r_{\rm in}$ (disc inner edge) and $r_{\rm out}$ (disc outer edge) in black dotted lines and the associated $1\sigma$ intervals as grey-shaded regions. Assuming that more than a few precession cycles have taken place (an assumption we examine below in \S\ref{sec:timescales}), both discs could be warped as $I_f$ changes smoothly as a function of $a$. Specifically, in HD 113337, the inner portion of the disc would trace the planet's orbit while the outer portion would be more closely aligned with the stellar companion's orbit. For HD~38529, we expect the disc to be more closely aligned with the outer planet's orbit, although the outer edge of the disc could be misaligned depending on the true separation of the stellar companion.

\subsubsection{Precession time-scales}\label{sec:timescales}

\begin{figure*}
\centering
\begin{subfigure}
    \centering
    \includegraphics[width=0.48\linewidth]{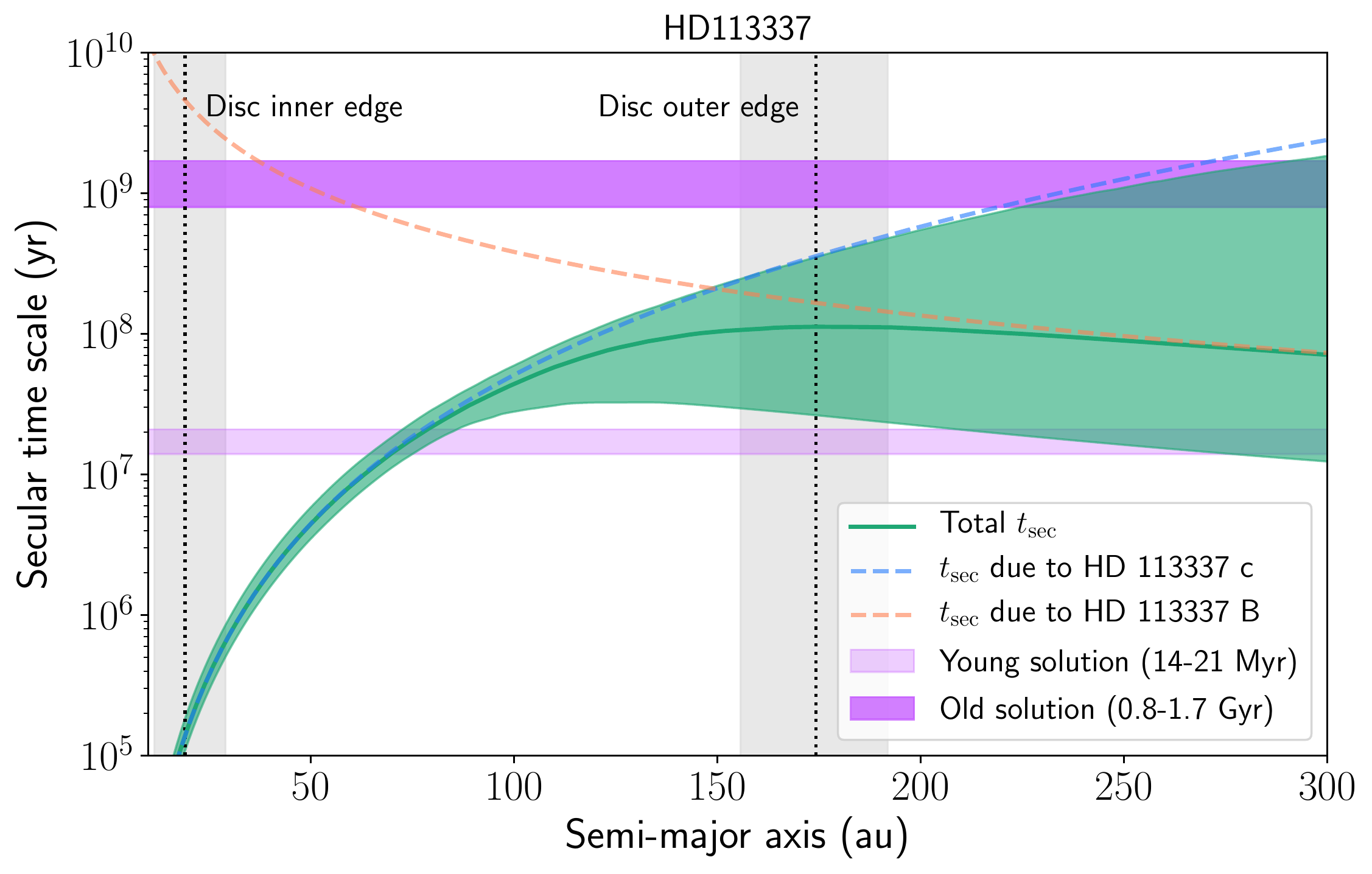}
    \hspace{2mm}
\end{subfigure}
\begin{subfigure}
  \centering
  \includegraphics[width=0.48\linewidth]{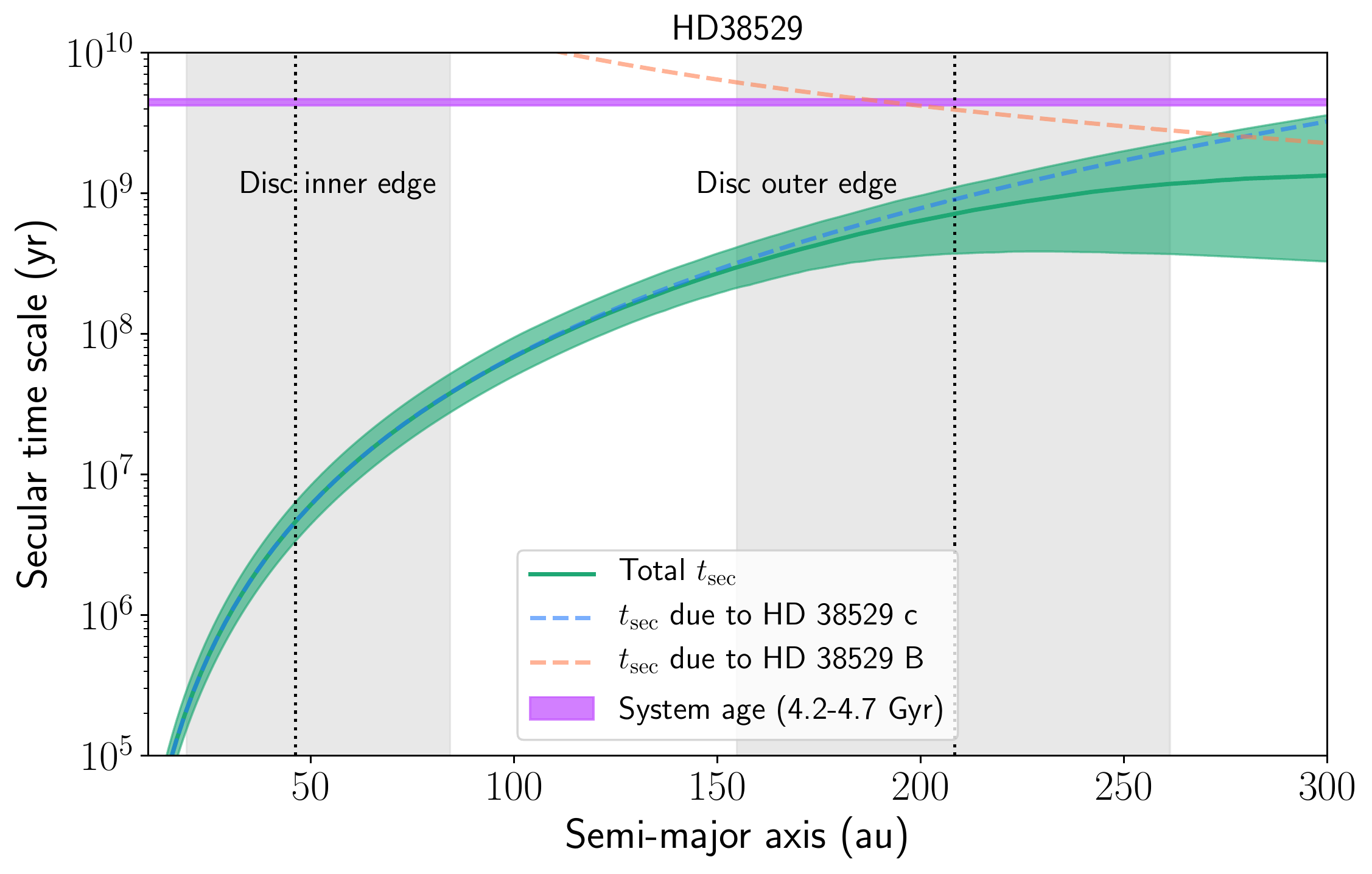}
\end{subfigure}
    \caption{The secular precession time-scale ($t_{\rm sec}$) of planetesimals as a function of semimajor axis under the influence of the outer planets and stellar companions for HD~113337 (left) and HD~38529 (right). The median $t_{\rm sec}$ is traced by the solid green line, while shaded regions around the line denote $2\sigma$ intervals created by taking into account uncertainties in $a$ and $m$ of the planet and binary. The dashed blue and orange lines show the median $t_{\rm sec}$ associated with only the planet (blue) and only the binary (orange), respectively. For both systems, $1\sigma$ intervals for the disc inner ($r_{\rm in}$) and outer ($r_{\rm out}$) edges are shown as grey regions, and median $r_{\rm in}$ and $r_{\rm out}$ values are plotted as black dotted lines. For HD~113337, the two possible age solutions from \citet{borgniet_constraints_2019} are shown as light and dark purple bars. For HD~38529, the age estimate from \citet{henry_host_2013} is shown as a dark purple bar.}
    \label{fig:sec_t}
\end{figure*}

To assess whether secular precession of the disc occurs within the system age for HD 113337 and HD 38529, we estimate the precession time-scale in this section. The precession rates are given by Eq.~\ref{eq:Bj}, so the total precession time-scale is
\begin{equation}\label{eq:total_tsec}
    t_{\rm{sec}}=2\pi/ |B_1 + B_2|.
\end{equation}
In Fig.~\ref{fig:sec_t}, we plot the median $t_{\rm sec}$ for HD~113337 (left panel) and HD~38529 (right panel) as green solid lines, with the shaded regions around the lines denoting the $2\sigma$ intervals for $t_{\rm{sec}}$. We also plot the median $t_{\rm sec}$ due solely to the outer planet and solely to the stellar companion in dashed blue and orange lines, respectively, to demonstrate which perturber dominates at a given distance. The disc extent is shown by plotting the median (black dotted lines) and $1\sigma$ intervals (grey-shaded regions) for $r_{\rm in}$ and $r_{\rm out}$. Finally, for HD~113337, the two possible age solutions from \citet{borgniet_constraints_2019} are plotted in light purple (14-21 Myr) and dark purple (0.8-1.7 Gyr). For HD~38529, the age solution from \citet{henry_host_2013} is shown in dark purple (4.2-4.7 Gyr).

We see that $t_{\rm sec}$ is much shorter than the system age for HD~38529. For HD~113337, $t_{\rm sec}$ is also shorter than the system age if the old age solution is correct. If HD~113337 is young instead, secular perturbations can only influence the inner portion of the disc out to $\sim 70$ au. For both systems, the perturbations are mostly dominated by the outer planets. Specifically, HD~38529~B induces a similar $t_{\rm sec}$ as HD~38529~c only at the outer edge of the disc. On the other hand, HD~113337~B causes disc particles to precess on a similar time-scale as the outer planet does at $a\sim150$ au, and becomes the dominant perturber beyond that distance.

\subsection{Implications for the planet-disc mutual inclination}\label{sec:sec_implications}
Here, we summarize the most likely explanations for our observed planet-disc $\Delta I$ of $\sim25-30\degr$ (assuming the most aligned solutions), given the analysis above. 

We find that if HD~113337 is old (0.8-1.7 Gyr), its disc could be warped due to joint perturbations from HD~113337~c and HD~113337~B, as shown by the forced inclination plot (Fig.~\ref{fig:forcedI}, left panel). Such a warped disc may appear to have a mid-plane that is misaligned with the planet's orbit, as explained by Fig.~\ref{fig:complexI}. As the resolution of the {\it Herschel} images is low, the inferred disc mid-plane from the images would be dominated by the outer disc, which is more resolved than the inner disc. Therefore, if the outer disc is preferentially aligned with the stellar companion as in Fig.~\ref{fig:forcedI}, {\it Herschel} images would preferentially trace the orientation of the outer disc, which is more misaligned with the planet than the inner disc is. This would act to amplify the observed planet-disc $\Delta I$, so we expect that higher-resolution disc imaging would show a mid-plane at the inner edge of the disc. 

On the other hand, if HD~113337 is young (14-21 Myr), secular perturbations can only induce precession for a portion the disc out to $\sim 70$ au, which in this case tends to align the disc to the orbit of HD~113337~c (as $I_f$ follows the planet inclination near the disc inner edge). The majority of the disc beyond $\sim 70$ au could have been misaligned with HD~113337~c from early on, for example in the protoplanetary disc phase. Due to the same resolution issue, a misaligned outer disc would dominate the disc orientation as seen from {\it Herschel}, which may explain the observed $\Delta I\sim25\degr$ for HD~113337, assuming the system had a protoplanetary disc that was misaligned with HD~113337~c.

The HD~38529 disc could also be warped (Fig.~\ref{fig:forcedI}, right panel), although to a lesser extent than the HD~113337 disc. If HD~38529~B has a small true semimajor axis, so the true $I_f$ is near the top of the orange region in Fig.~\ref{fig:forcedI}, the disc might be warped enough to produce some amount of planet-disc $\Delta I$, which is enhanced because {\it Herschel} resolves the outer disc better than the inner disc. Alternatively, HD~38529~B may be misaligned by more than $25\degr$ with HD~38529~c, which could enhance the resulting planet-disc $\Delta I$. 

However, if either HD~38529~B or HD~113337~B is misaligned by more than $\sim40\degr$ with its disc (which we assume is coplanar with the outer planet), it could induce Kozai-Lidov oscillations and drastically re-shape the disc so that the dynamics can no longer be described with Laplace-Lagrange theory. If the perturbing body is highly misaligned, the resultant disc would also become nearly spherical and therefore look circular \citep{verrier_hd_2008,farago_high_2010,kennedy_99_2012,pearce_dynamical_2014}. This is likely not consistent with the {\it Herschel} images for at least the HD~38529 disc, which is better described by a thin disc that is viewed inclined, although it is harder to rule this out for the nearly face-on HD~113337 disc. We note that these considerations also argue against the larger $\Delta I$ solutions (the second prograde solutions in \S\ref{sec:deltaI}), as those could require even larger values of planet-binary $\Delta I$.

From an observational standpoint, the inferred mid-plane and hence inclination of a warped disc would depend on many factors, including the surface density profile, temperature profile, and the relative geometry between different segments of the disc, each of which have a different forced inclination. As mentioned above, the low resolution of {\it Herschel} images also means that the inferred disc mid-plane will be dominated by the outer parts of the disc, which are more resolved. To zeroth order, we expect the disc to have an average mid-plane that is between the minimum and maximum $I_f$ in Fig.~\ref{fig:forcedI}. Therefore, for both systems, the outer planets and debris discs might acquire $\sim10\degr$ misalignments after a few cycles of precession given planet-binary $\Delta I$ of $25\degr$ (and if HD~113337 is old). This conclusion does not depend on the initial disc inclination, as $I_f$ is independent of whether the planet and disc are originally aligned, which we assumed for simplicity in Fig.~\ref{fig:complexI}. $\Delta I\sim 10\degr$ is slightly lower than the best fit planet-disc $\Delta I$ and $1\sigma$ regions for both systems, assuming the most aligned solutions (see Fig.~\ref{fig:Imut}), but is consistent with these solutions at the $2\sigma$ level (see \S\ref{sec:deltaI}).

If the discs are misaligned with the outer planets because their outer regions are torqued by the stellar companions, they would also become puffed up or vertically extended, as illustrated by Fig.~\ref{fig:complexI}. For example, for HD~113337, if the disc and HD~113337~c were initially aligned, and $I_f$ is between 0 and $\sim20\degr$ (Fig.~\ref{fig:forcedI}), after precession the disc could become vertically extended over an angular region of $\sim40\degr$ at its outer edge. High-resolution imaging of the debris discs in HD~113337 and HD~38529 could be used to test whether the discs are warped and constrain their vertical scale heights.

Finally, it is possible that the stellar companions are farther out and the true $I_f$ curves are toward the bottom of the $2\sigma$ regions in Fig.~\ref{fig:forcedI}, in which case they cannot be responsible for the observed planet-disc misalignment. We discuss future observations which could better constrain the planetary orbits and disc orientations in \S\ref{sec:discuss}.

\subsection{Orientation of the stellar spin axis}\label{sec:stellar_spin}
To further understand the system architecture of HD~113337 and HD~38529, it is helpful to know the relative orientation of the stellar spin axis with respect to the planetary orbit and debris disc. For example, the spin axis of $\beta$ Pic is found to be aligned with the orbit of its outer planet $\beta$ Pic b \citep{kraus_spin-orbit_2020-1}. The orbit of $\beta$ Pic b is also closely aligned with its disc (planet-disc $\Delta I\approx2.4\degr$, \citealt{matra_kuiper_2019}), and the orbit of $\beta$ Pic c (less than a degree, \citealt{nowak_direct_2020}), suggesting that the planets formed in a coplanar disc without significant primordial misalignments. In addition, \citet{greaves_alignment_2014} compared disc inclinations with inclinations of the stellar spin axis (using $v \sin{I_\star}$ and $P_\star$) for 11 systems, and found that all systems are consistent with star-disc $\Delta I\lesssim10\degr$, suggesting the lack of any substantial misalignments that could be produced by misaligned perturbers (with the caveat that differing ascending nodes could cause undetected misalignments). On the other hand, \citet{davies_star-disc_2019-1} compared the inclinations of the stellar spin axis and protoplanetary discs and found that four out of ten systems with good $I_\star$ estimates show potential misalignments.

For HD~113337, we estimate that $I_\star=24^{+30}_{-10}\degr$ (see \S\ref{sec:113337_intro}). In comparison, $I_c=31^{+5}_{-4}\degr$ for the outer planet, and $I_{\rm{disc}}=13^{+10}_{-9}\degr$. For HD~38529, we estimate that $I_\star=69\pm14\degr$ (see \S\ref{sec:38529_intro}), which is equivalent to $I_\star=111\pm14\degr$. The outer planet in HD~38529 has $I_c=135^{+8}_{-14}\degr$, and the disc has $I_{\rm{disc}}=71^{+10}_{-7}\degr$, or equivalently, $I_{\rm{disc}}=109^{+10}_{-7}\degr$. Therefore, in both systems, the star is consistent with being aligned with or nearly aligned with the outer planet's orbit and the debris disc. The exact mutual inclinations will depend on the line of nodes of the stellar rotation, but the fact that the three inclinations could lie close to each other is suggestive of system-wide alignment, or at least near alignment. This also supports the most aligned $\Delta I$ solutions as being the real ones for both systems, rather than the other solutions which indicate larger misalignments or retrograde orbits.

\section{Discussion}\label{sec:discuss}

\subsection{Overview of results}
In summary, we measured the planet-disc mutual inclinations ($\Delta I$) between the long-period giant planets and debris discs in HD~113337 and HD~38529. To do this, we first constrained the 3-D orbit of the outer giant planets (HD~113337~c and HD~38529 c) by combining {\it Gaia} DR2-{\it Hipparcos} astrometry with RV data. Then, we modelled the {\it Herschel} resolved images of the debris discs to get the disc inclination and position angle. Although the nature of disc measurements leaves four degenerate disc orientations, we find that the orientations which give the most aligned or minimum $\Delta I$ solutions are likely the real ones for both systems.

From our distributions for the planet-disc $\Delta I$, the most aligned solutions give median values of $\Delta I\sim25-30\degr$. Specifically, we find $\Delta I=17-32$ ($1\sigma$), $\Delta I =9-40$ ($2\sigma$), and $\Delta I=3-50\degr$ ($3\sigma$) for HD~113337. For HD~38529, we find $\Delta I=21-45\degr$ ($1\sigma$), $\Delta I=10-116\degr$ ($2\sigma$), and $\Delta I=3-144\degr$ ($3\sigma$). For both systems, the planets and discs are therefore inconsistent with being aligned at $\gtrsim3\sigma$ confidence. In \S\ref{sec:assess_align}, we discussed the dynamics of the disc by considering the influence of both the outer planets and stellar companions, and estimated the inclination of the stellar spin axis in HD~113337 and HD~38529 to see whether it is aligned with the planet and disc. In the next two subsections, we summarize the evidence for planet-disc misalignment (\S\ref{sec:case_align}), and arguments for alignment (\S\ref{sec:case_misalign}). Finally, we discuss in \S\ref{sec:future} future observations that could better decide between misalignment and alignment.

\subsection{The case for misalignment}\label{sec:case_misalign}
In \S\ref{sec:sec_evol}, we find the debris discs in HD~38529 and HD~113337 could have forced inclinations that lie between the orbits of the outer planets and stellar companions, assuming their orbits are misaligned. As a result, precession of the disc particles around the forced inclination plane causes the disc mid-plane to be misaligned with the planet's orbit, even if the disc was initially aligned with the planet. In fact, the discs would also become warped as the forced inclination is different for disc particles at different distances (e.g. it is closer to the planet near the planet's orbit). For both systems, we find that the disc precession time-scales induced by the inner planets are much longer than that of the outer planets, and ignore them in our analysis (the 3-D orbits of the inner planets are also unknown).

HD~113337~B has a closer projected separation ($\sim4000$ au) than HD~38529~B ($\sim11000$ au), and more effectively competes with its outer planet in torquing the disc (see Fig.~\ref{fig:forcedI}). If HD~113337 is old, then joint perturbations from HD~113337~c and HD~113337~B provide a natural way to explain our observed planet-disc misalignment, as the secular precession time-scale is lower than the old age solution (0.8-1.7 Gyr). To reach planet-disc $\Delta I\sim25\degr$, the disc of HD~113337 could be puffed up by as large as $\sim25\degr$ in angular scale height, making the disc close to spherical and therefore appear circular to the observer. This cannot be ruled out by our {\it Herschel} images, given that HD~113337's disc is inferred to be a thin disc seen nearly face-on (which is degenerate with an almost spherical disc). The planet-disc $\Delta I$ could be as small as $\sim9\degr$ at $2\sigma$, however, so the disc may not need to be so puffed up. On the other hand, if HD~113337 is young, then neither the planet nor the stellar companion would have enough time to significantly influence the disc, so the observed misalignment would be primordial, possibly arising during the protoplanetary disc phase.

HD~38529~B is not as effective in perturbing the disc, but due to large uncertainties in its orbit, it may still induce a forced inclination that causes the outer planet and disc mid-plane to be misaligned by $\sim10\degr$ (see Fig.~\ref{fig:forcedI}). The age of HD~38529 is $4.2-4.7$ Gyr, so there is sufficient time for secular precession to re-shape the disc. A $\sim10\degr$ misalignment would again be consistent with our $\Delta I$ measurements for HD~38529 at the $2\sigma$ level. 

For HD~38529 c, independent orbit measurements using HST \citep{benedict_mass_2010} find a 3-D orbit consistent with ours, after adjusting their orbital angles to have the correct orbital direction (see \S\ref{sec:orbit_results} for details). Specifically, we measured $I_c = 135^{+8}_{-14}\degr$, $\Omega_c = 217^{+15}_{-19}\degr$ using {\it Gaia}-{\it Hipparcos} astrometry, while the \citet{benedict_mass_2010} HST measurement (after adjustment) is $I_c = 131\pm4\degr$ and $\Omega_c = 218\pm8\degr$. In fact, the lower error bars of the \citet{benedict_mass_2010} measurement would translate to even stronger evidence for planet-disc misalignment in HD~38529, which renders near alignment more unlikely.

\subsection{The case for alignment}\label{sec:case_align}
Despite the above discussion, it is however possible that the outer planet and debris disc are aligned in HD~38529 or HD~113337, or both. Our measurements of the planet-disc $\Delta I$ give a small probability for near alignment ($<0.3$ per cent chance for $\Delta I\lesssim3\degr$). As pointed out in \S\ref{sec:stellar_spin}, the fact that the stellar spin inclinations are consistent with the planet and disc inclinations in both systems argues for alignment or near alignment between the planet and disc. Previous studies have found several systems consistent with alignment, including $\beta$ Pic, HR 8799, HD 82943, and AU Mic. If star-planet-disc alignment is also the case for HD~113337 and HD~38529, this could provide evidence that planetary systems forming in protoplanetary discs are typically coplanar, partly due to the protoplanetary discs themselves being flat. Alternatively, if large scale heights are observed for the discs, that could indicate the discs were primordially misaligned with the planets but later re-aligned (and that the stellar companions are too far away to influence the discs).




\subsection{Future constraints}\label{sec:future}

Improved observations of the debris discs and planetary orbits would provide stronger conclusions about the (mis)alignment status of these two systems. For instance, {\it Gaia} epoch astrometry is expected to reach a precision of $\sim10~\mu \rm{as}$ and detect the astrometric wobble of $\sim20000$ stars as caused by orbiting long-period giant planets \citep{perryman_astrometric_2014}. Given their parallaxes, semimajor axes, and planet-host star mass ratios, the astrometric signature from HD~113337~c and HD~38529~c would both have amplitudes of $\gtrsim1$ mas, so would be easily detectable by {\it Gaia}. The periods of these companions are also relatively short ($\sim6$ yr for HD~38529~c and $\sim9$ yr for HD~113337 c), meaning that a five-year baseline for {\it Gaia} epoch astrometry would cover a large fraction of the orbits. These measurements, combined with the RV data and proper motion anomaly constraints, would significantly reduce the uncertainties in the planet's orbital inclination and ascending node.

Imaging the debris discs at higher resolution would reduce uncertainties for the disc orientation, especially for the nearly face-on disc of HD~113337, which has a highly unconstrained position angle from {\it Herschel} images. Indeed, high-resolution disc imaging could directly probe whether or not the discs are warped, or measure their vertical scale heights \citep[e.g.][]{matra_kuiper_2019, daley_mass_2019}. If the discs are found to be significantly warped and have large vertical scale heights, that would provide strong evidence that the planet-disc misalignments are caused by joint perturbations from the outer planets and stellar companions in these systems (see \S\ref{sec:sec_evol} and \S\ref{sec:sec_implications}).

In the meantime, a better age constraint for HD~113337 would be helpful. If the young age solution for HD~113337 is confirmed, the misalignment we observe between HD~113337~c and its debris disc could be primordial, making this system an interesting target for high-resolution disc observations, and a valuable testbed for studying early planet-disc misalignments. Indeed, theoretical studies have found that internal inclined giant planets are able to induce observable misalignments and warps in protoplanetary discs \citep{zhu_inclined_2019,nealon_scattered_2019}. Given our relatively well constrained orbit of HD~113337 c, a young system age would also make the planet a promising target for future direct imaging instruments, although the separation could be challenging ($\sim120$ mas).

\section*{Acknowledgements}
We thank the anonymous referee for a positive report. JX is grateful to Simon Borgniet for sharing the SOPHIE RV data for HD~113337. JX acknowledges funding from the Downing Scholarship of Pomona College and Downing College, Cambridge. GMK is supported by the Royal Society as a Royal Society University Research Fellow. This work has made use of data from the European Space Agency (ESA) mission {\it Gaia} (\url{https://www.cosmos.esa.int/gaia}), processed by the {\it Gaia} Data Processing and Analysis Consortium (DPAC, \url{https://www.cosmos.esa.int/web/gaia/dpac/consortium}). Funding for the DPAC has been provided by national institutions, in particular the institutions participating in the {\it Gaia} Multilateral Agreement. {\it Herschel} is an ESA space observatory with science instruments provided by European-led Principal Investigator consortia and with important participation from NASA.

\section*{Data Availability}
The proper motion anomaly data used in this paper are available in \citep{brandt_erratum_2019}, at \url{https://doi.org/10.3847/1538-4365/ab13b2}. The RV data for HD~38529 are available in \citet{tal-or_correcting_2019} at \url{http://vizier.u-strasbg.fr/viz-bin/VizieR?-source=J/MNRAS/484/L8}, and the RV data for HD~113337 are available in \citet{borgniet_extrasolar_2019} on reasonable request to the corresponding author of that paper.

\bibliographystyle{mnras}
\bibliography{discs} 

\appendix

\section{Orbit posteriors of HD~38529~B}\label{appendixA}

\begin{figure}
    \centering
    \includegraphics[width=0.8\linewidth]{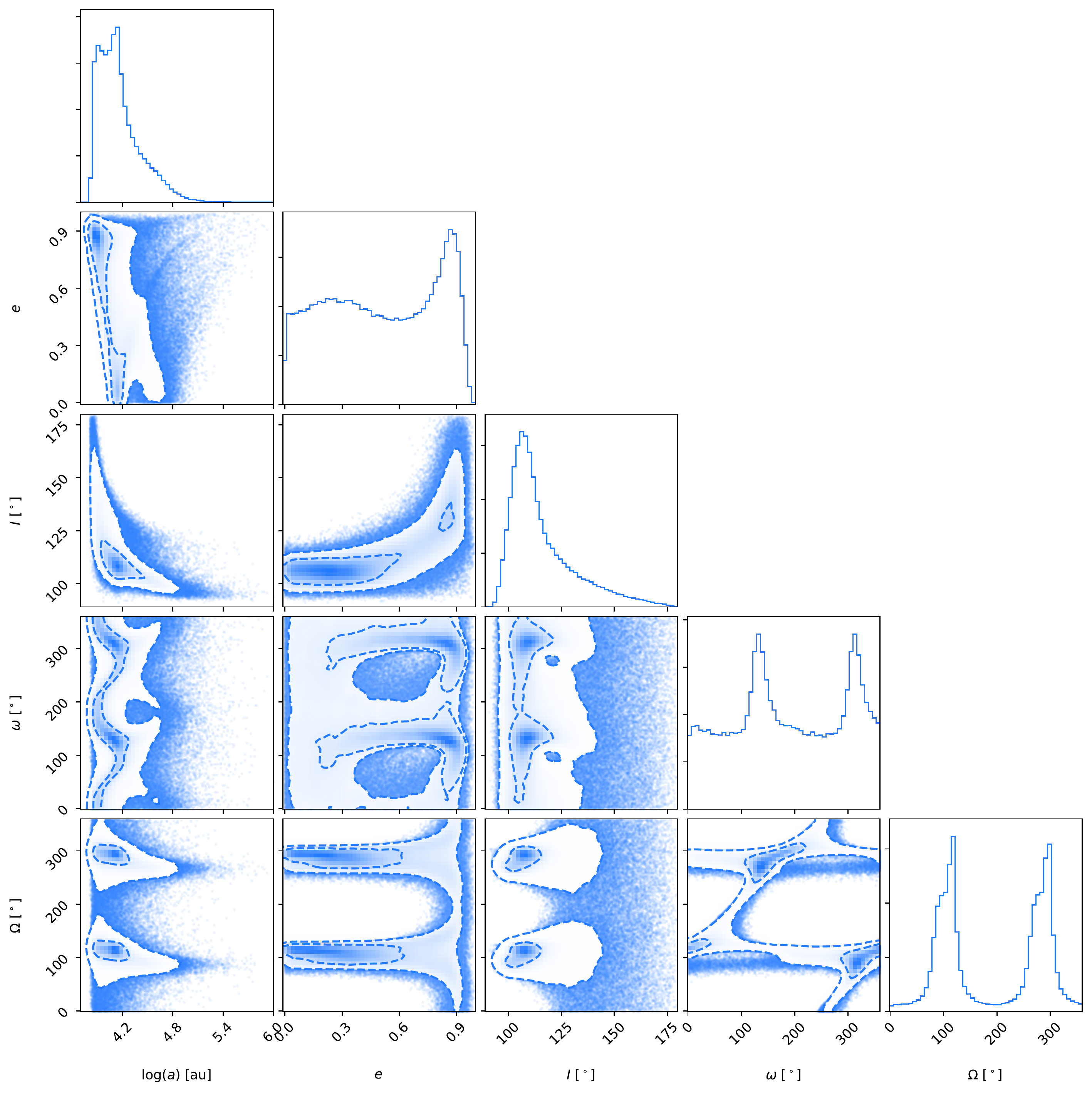}
    \caption{Joint posterior distribution of orbital elements for HD~38529 B generated with relative {\it Gaia} DR2 measurements, showing the logarithmic semimajor axis log($a$), eccentricity $e$, inclination $I$, argument of periastron $\omega$, and longitude of ascending node $\Omega$.}
    \label{fig:38529B_orbit}
\end{figure}

\section{Full posterior distributions for outer planets}\label{appendixB}

\begin{figure*}
    \centering
    \includegraphics[width=1.0\linewidth]{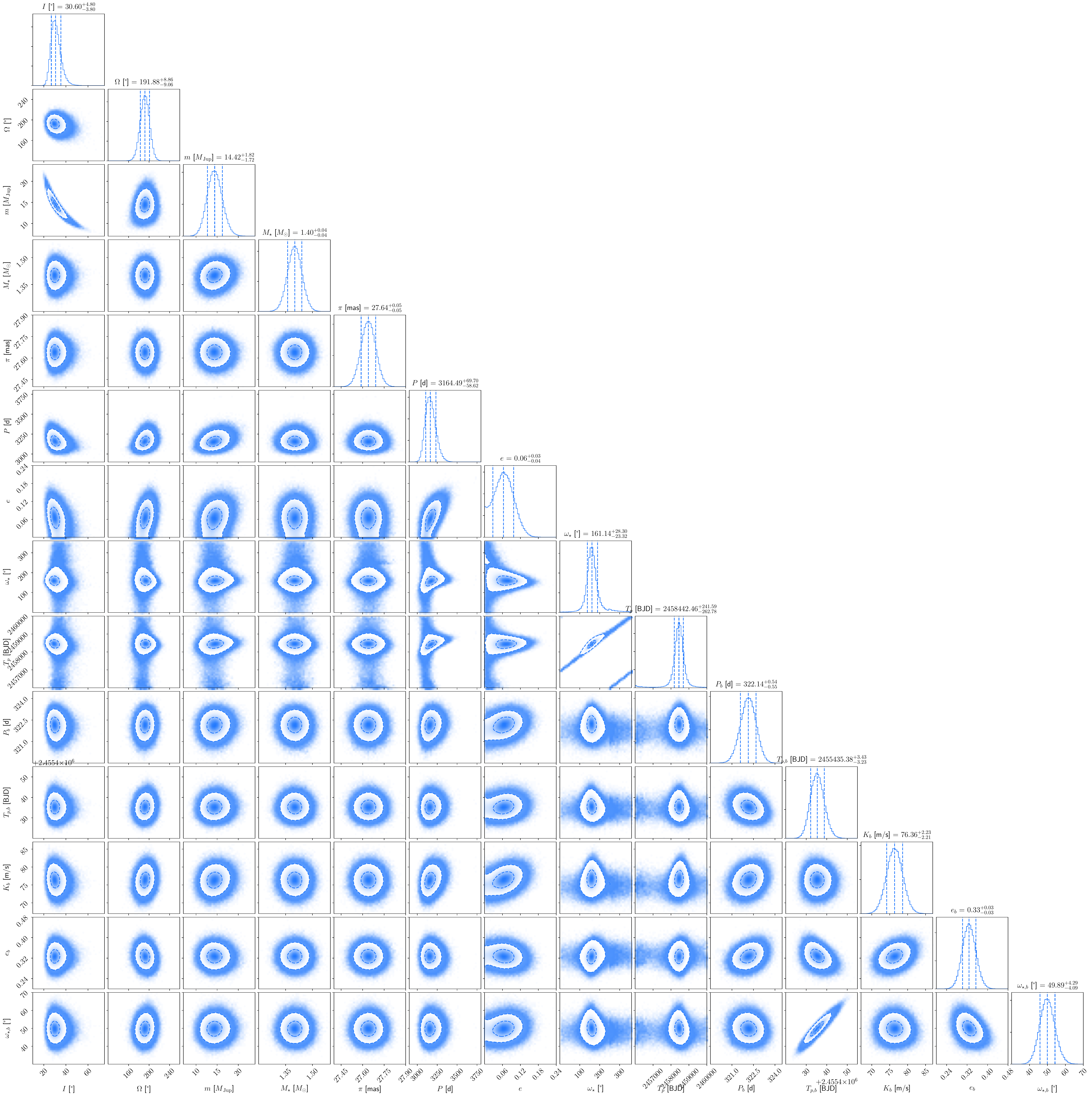}
    \caption{Target: HD~113337. Joint posterior distributions for the orbital parameters of HD~113337~c and b from our joint RV and PMa fits. The planet parameters without subscripts are for planet c. Moving outward, the dashed lines on the 2D histograms correspond to $1\sigma$ and $2\sigma$ contours.}
    \label{fig:113337_fullcor}
\end{figure*}

\begin{figure*}
    \centering
    \includegraphics[width=1.0\linewidth]{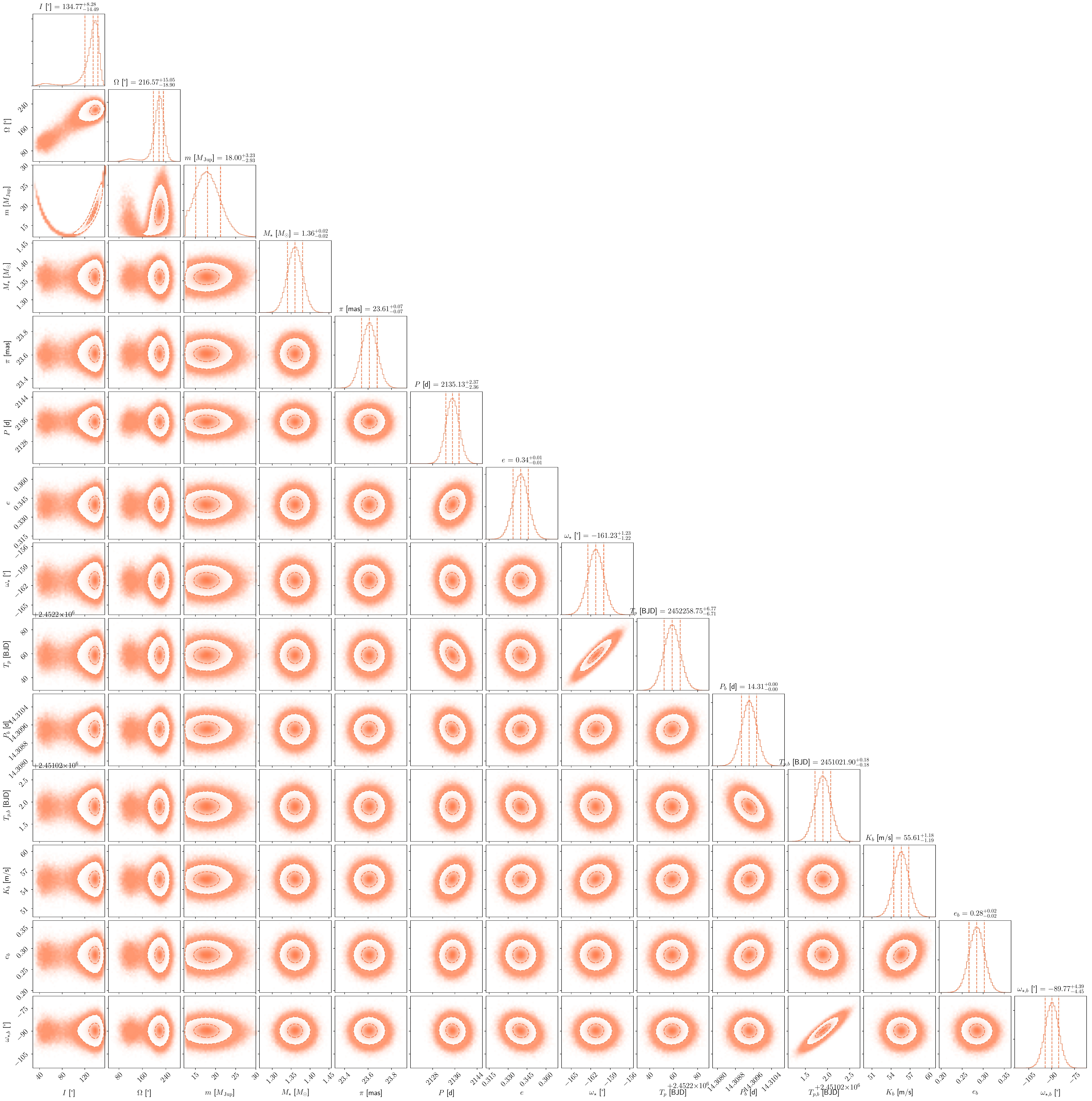}
    \caption{Target: HD~38529. Joint posterior distributions for the orbital parameters of HD~38529~c and b from our joint RV and PMa fits. The planet parameters without subscripts are for planet c. Moving outward, the dashed lines on the 2D histograms correspond to $1\sigma$ and $2\sigma$ contours.}
    \label{fig:38529_fullcor}
\end{figure*}

\bsp	
\label{lastpage}
\end{document}